\documentclass[preprint2]{aastex62}
\usepackage[utf8]{inputenc}
\usepackage{mathtools}
\usepackage{graphicx} 

\begin{document}

\title{Thermal Sunyaev-Zel'dovich Measurements of Locally Bright Galaxies with ACT DR6: Radio Source Contamination and Excess Compton-$y$ Signal}

\author{Nicholas Battaglia}
\affiliation{Department of Astronomy, Cornell University, Ithaca, NY 14853, USA}
\affiliation{Universit\'{e} Paris Cit\'{e}, CNRS, Astroparticule et Cosmologie, F-75013 Paris, France}
\affiliation{Universi\'{e} Paris-Saclay, CNRS, Institut d'Astrophysique Spatiale, 91405, Orsay, France}

\author{Jean-Baptiste Melin}
\affiliation{Universit\'{e} Paris-Saclay, CEA, D\'{e}partement de Physique des Particules, 91191, Gif-sur-Yvette, France}

\author{J.~Colin Hill}
\affiliation{Department of Physics, Columbia University, New York, NY 10027, USA}

\author{James Bartlett}
\affiliation{Universit\'{e} Paris Cit\'{e}, CNRS, Astroparticule et Cosmologie, F-75013 Paris, France}

\begin{abstract}
The Planck collaboration found a remarkable power-law relation between stellar mass and the thermal Sunyaev-Zeldovich (tSZ) signal for the Locally Bright Galaxy (LBG) sample, spanning over a decade in stellar mass. We re-examine this measurement using the Atacama Cosmology Telescope (ACT) DR6 component-separated Compton-$y$ maps, which provide lower noise and higher angular resolution than Planck, on a footprint spanning one-third of the sky. We recover a consistent power-law scaling between the cylindrical Compton-$y$ signal and stellar mass. Additionally, we identify residual contamination in the tSZ signal from radio sources at the few percent-level, which has not been considered previously. In parallel, we identify a factor-of-two excess in the Compton-$y$ signal in LBGs hosting co-spatial radio sources relative to those without, at fixed stellar mass. This excess persists to radii of at least 6 arcminutes, suggesting a halo-scale effect, and is recovered in the original Planck results when the same radio source subselection is applied. We consider two physical explanations: a systematic difference in halo mass at fixed stellar mass, or thermal energy injected into the circumgalactic medium by Active Galactic Nuclei, although we cannot currently distinguish between the two. This result has direct implications for tSZ cross-correlation measurements more broadly, and necessitates careful characterization of the radio source fraction in galaxy samples in future analyses.

\end{abstract}

\section{Introduction}

The connection between a galaxy's mass and its gas properties is fundamental to our understanding of how galaxies evolve across cosmic time. The circumgalactic medium (CGM) contains the majority of baryons, mostly ionized, within a galaxy and roughly spans from the outer boundary of the interstellar medium (ISM) to the intergalactic medium (IGM). The properties of the CGM play a crucial role in a galaxy's {\it cosmic ecosystem}, a central theme of the 2020 Decadal Survey \citep{Astro2020}, and to zeroth order the CGM's thermodynamic properties should largely depend on the galaxy's halo mass \citep[e.g.,][]{Planck2013}. However, the interplay between the CGM and various feedback processes, including stellar winds, supernovae, and active galactic nuclei (AGN), combined with radiative cooling, will additionally shape the thermodynamic properties of the CGM and govern how gas accretes onto the galaxy, where it may eventually cool and form stars.

There are many ways to probe the CGM, such as absorption line studies \citep[e.g.,][]{Tumlinson2017} and X-ray emission \citep[e.g.,][]{Erosita2022,Erosita2024}. Here we focus on using the cosmic microwave background (CMB) as a backlight to probe the outer CGM, defined here as regions beyond 150\,kpc, as well as the transition zone from the outer CGM into the IGM surrounding the galaxies. As CMB photons travel through the expanding Universe, they scatter off energetic electrons in the intracluster medium (ICM) and CGM, creating secondary fluctuations that dominate temperature anisotropies on arcminute scales \citep{1970Ap&SS...7....3S}. The thermal SZ effect (tSZ) is the increase in the energy of CMB photons due to inverse Compton scattering off hot electrons, allowing us to directly trace the line-of-sight pressure of ionized baryons in the CGM. The distortion of the CMB blackbody spectrum by the tSZ effect is
\begin{equation} \label{eq:TtSZ}
    \frac{\Delta T(\nu)}{T_{\text{CMB}}} = f(\nu) \frac{\sigma_T}{m_e c^2} 
    \int_{\text{LOS}} P_e \mathrm{d}l \ ,
\end{equation}
where $\Delta T(\nu)$ is the temperature deviation caused by the tSZ effect at frequency $\nu$, $T_{\text{CMB}} = 2.725$\,K is the CMB temperature, $f(\nu) = x\coth(x/2) - 4$ is the spectral function with $x = h\nu / k_B T_{\text{CMB}}$, and $h$ and $k_B$ are the Planck and Boltzmann constants, respectively. Relativistic corrections \citep[e.g.,][]{Itoh1998,Chluba2012} are neglected here, as the gas temperatures in galaxies are appreciably lower than in galaxy clusters. The frequency-independent amplitude of 
Equation~\ref{eq:TtSZ} is the Compton-$y$ parameter,
\begin{equation} \label{eq:y}
    y = \frac{\sigma_T}{m_e c^2} \int_{\text{LOS}} P_e \mathrm{d}l.
\end{equation}

There are several measurements of the average tSZ signal from large samples of galaxies \citep[e.g.,][]{Hand2011,Spacek2016,Koukoufilippas2020,Meinke2021,Schaan2021,Sanchez2023,Liu2025}. The Planck collaboration \citep{Planck2013} measured the tSZ signal from the Locally Bright Galaxy (LBG) sample and found a remarkably tight power-law relationship between the total mass of the galaxy and the tSZ signal over a decade in mass, independently confirmed by \citet{Greco2015} in the high signal-to-noise regime. More recently, several tSZ measurements have pushed to provide spatially resolved information by measuring tSZ radial profiles. Such measurements, extending to and beyond the virial radius, are complementary to traditional inner CGM observations and provide information on the thermodynamic profiles and total thermal energy of the CGM\citep{BFSS2017}.

Constraining the thermal energy of the CGM provides a key window into its energetics, as the CGM is expected to have additional heating mechanisms beyond gravitational collapse. Energy injected by active galactic nuclei (AGN feedback) and supernovae will heat the CGM, where cooling times in the outer regions are long, on the order of Gyr. Cosmological simulations have shown that in massive halos AGN feedback measurably alters the tSZ signal in massive halos \citep[e.g.,]{Scannapieco2008,BBPSS,McCarthy2014,Grayson2023}. Previous cross-correlations with AGN samples have also detected tSZ signals \citep{Crichton2016,Verdier2016,Soergel2017,Hall2019}, and from these measurements it has been possible to estimate the tSZ contribution from the AGN, under assumptions about the host galaxy mass and its intrinsic signal. Having a control sample of galaxies at the same masses that do not host AGN would remove the need for such assumptions, enabling a strictly empirical comparison.

Spatially resolving the thermal energy as a function of radius out to and beyond the virial radius pushes observational requirements to high-resolution tSZ measurements across a range of galaxy samples including higher redshifts. Such measurements are more readily achievable with ground-based CMB experiments, such as the Atacama Cosmology Telescope (ACT) and the South Pole Telescope (SPT), than with Planck, owing to their higher angular resolution. The larger beam of Planck dilutes the tSZ signal on CGM scales, limiting its ability to spatially resolve the pressure profile and disentangle the tSZ signal from other contributions. For tSZ cross-correlations with galaxies, contaminants such as cosmic infrared background (CIB) emission from dusty star-forming galaxies and radio emission from AGN are expected, as both are found in galaxies. When isolating the tSZ signal via component separation, residual CIB emission largely arises from variations in CIB spectral energy distributions \citep[e.g.,][]{Raghunathan2023,McCarthyHill2024}. Contamination from radio point sources is expected to be smaller, since 90\,GHz is relatively high for bright synchrotron emission not already removed in CMB maps. Recently, high-resolution observations with MUSTANG-2 on the Green Bank Telescope found numerous radio point sources in galaxy clusters selected via their tSZ signal by ACT \citep{Dicker2021,Dicker2024}; while these sources are predominantly faint, their presence in galaxies with lower tSZ flux can become an important residual systematic to characterize and mitigate. This potential contamination was pointed out in a previous Fourier-space, tSZ-lensing cross-correlation theoretical forecast \citep{Shirasaki2019}.

In this work we re-analyze the LBG sample from \citet{Planck2013} using the ACT Data Release 6 (DR6) component-separated Compton-$y$ maps \citep{Coulton24}, assess the presence of residual radio source contamination in tSZ cross-correlations, and measure the tSZ signal for subsamples of LBGs with and without co-spatial radio sources. This subdivision effectively creates a control sample of galaxies with and without a radio source (most likely radio-loud AGN) at fixed stellar mass, enabling an empirical test of the impact of AGN feedback on the CGM. We organize this paper as follows: Section~\ref{sec:data} describes the ACT DR6 data and the LBG sample; Section~\ref{sec:meth} details the compensated aperture photometry (CAP) flux extraction method and stacking procedure; Section~\ref{sec:results} presents the Compton-$y$ measurements as a function of stellar mass and the radial profiles for the two subsamples; Section~\ref{sec:disc} discusses the implications for halo mass differences and AGN feedback energetics; and Section~\ref{sec:conc} summarizes our conclusions. Systematic tests are presented in Appendix~\ref{sec:sys}.

\begin{figure*}[ht!]
\label{fig:gallery}
\centering      
    \includegraphics[width=0.98\textwidth]{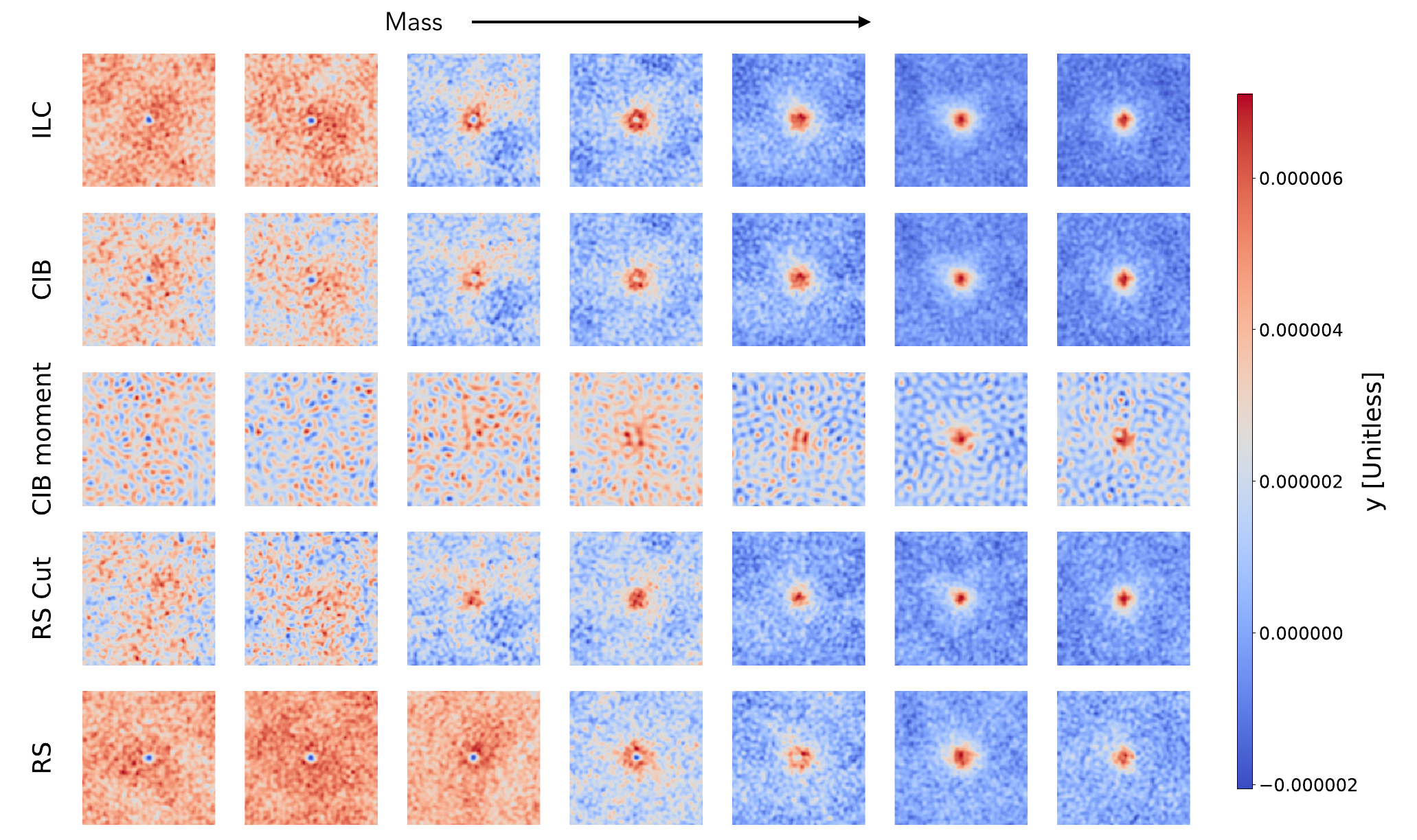}
\caption{Gallery of co-added, component-separated ACT(+Planck) Compton-$y$ maps on the overlapping LBG sample galaxies. Each row illustrates a co-add of a given component-separated map: The top row are the fiducial ACT ILC Compton-$y$ map co-adds (labeled ILC); the second row illustrates the de-projected CIB ILC Compton-$y$ map co-adds (labeled CIB); the third row illustrates the moment-deprojected CIB ILC Compton-$y$ map co-adds (labeled CIB moment); and the fourth and fifth rows illustrate the CIB ILC Compton-$y$-map co-adds that exclude co-located radio sources and include co-located radio sources (labeled RS cut and RS, respectively). For a given component-separated map, the columns subdivide the LBG sample into stellar mass bins described in Section~\ref{sec:LBG}, where the description for the radio source cuts is also found. The leftmost column has a stellar mass range of $\mathrm{Log}_{10}(M_\star/M_\odot) = 11.1$ to $\mathrm{log}_{10}(M_\star/M_\odot) = 11.2$ and the rightmost column has a stellar mass range of $\mathrm{log}_{10}(M_\star/M_\odot) = 11.7$ to $\mathrm{log}_{10}(M_\star/M_\odot) = 11.8$, with the columns in between increasing from left to right by 0.1 in $\mathrm{log}_{10}(M_\star/M_\odot)$. The colorbar represents the dimensionless Compton-$y$ parameter and is common for all co-adds in this figure. Clearly, the noise level of the CIB moment deprojected co-adds is higher than the other co-adds. The negative source at the center of the co-adds is from co-located radio source contamination --- when the LBG sample excludes galaxies with co-located radio sources (RS Cut) the negative source is not visible (Row 4 versus Row 2).}
\end{figure*}

\section{Data}
\label{sec:data}

\subsection{ACT and Planck Data}

\label{sec:ILC}

We use the public DR6 component-separated maps from the ACT collaboration \citep{Coulton24}. These maps were made using the ACT data from the DR4 and DR6 releases \citep{Aiola2020,Naess25} in combination with data from the Planck satellite \citep{Planck2020}. When combining the individual frequency maps they are convolved to a common beam with a full width at half maximum (FWHM) of 1.6 arcmin.

The ACT DR6 component-separated maps were constructed using the internal linear combination (ILC) method \citep{Coulton24}. An ILC constructs a linear combination of multi-frequency maps by exploiting the known frequency dependence of a signal of interest \citep[e.g.,][]{Delabrouille2003}, such as the tSZ effect, and has been widely applied to CMB data \citep[e.g.,][]{PlanckYmap2016,PlanckYmap2020,Coulton24,Bleem2022}. A key advantage of the ILC is that it is a blind component separation technique; the frequency behavior of contaminants does not need to be known a priori. This is particularly important for the CIB, which, while broadly described as a modified blackbody, does not have a well-understood spectral energy distribution (SED) that can be derived from first principles. When the SED of a contaminant is known, this information can be incorporated into the ILC to construct a map that is completely insensitive to that contaminant \citep{Remazeilles2011}, at the cost of increased variance in the reconstruction. This constrained ILC, or deprojection, is especially useful in cross-correlation analyses where a contaminant is highly correlated with the signal of interest. In the case of tSZ cross-correlations with tracers that are more strongly correlated with the CIB than with the tSZ field itself, such as the CMB lensing potential, imperfect CIB removal due to an assumed SED can leave a residual bias in the measurement \citep[e.g.,][]{McCarthyHill2024}.

To mitigate this, a moment-based ILC approach was introduced to account for uncertainty in SED parameters \citep{Chluba2017,Remazeilles2021}. In this framework, the contaminant SED is expanded in a Taylor series around a pivot frequency, with each term, or moment, capturing increasingly higher-order spatial variation in the SED. By simultaneously deprojecting multiple moments, the ILC is made insensitive not only to the mean contaminant emission but also to its spectral variation across the sky. This is particularly relevant for the CIB, whose spatially varying dust temperatures mean a single fixed SED deprojection leaves residual bias; the moment approach suppresses this residual more robustly, at the cost of additional variance in the reconstruction. The ACT DR6 component-separated maps include three types of ILC methods applied to the maps on a needlet frame \citep{Remazeilles2011}, which use spherical wavelet transforms to provide additional spatial scale separation \citep[see][for more details on the application on ACT data]{Coulton24}. We show results using all three types of ILC methods to test the robustness of the results to the choice of ILC method.

\subsection{Locally Bright Galaxy Sample}
\label{sec:LBG}

The galaxy sample used in this work is the so-called Locally Bright Galaxy (LBG) sample, defined in \citet{Planck2013} with a similar sample obtained independently by \citet{Greco2015}. The galaxies in the LBG sample were selected to be nearby, bright, and predominantly central galaxies. The initial galaxy selection was drawn from the New York University Value Added Galaxy Catalog \citep[NYU-VAGC;][]{Blanton2005}, based on the seventh data release of the Sloan Digital Sky Survey \citep[SDSS DR7;][]{SDSSDR7}. Three additional criteria were then applied: a magnitude cut of $r < 17.7$ (r-band, extinction-corrected Petrosian magnitude), a redshift cut of $z > 0.03$, and a central galaxy selection. Central galaxies were defined as the brightest galaxies within 1.0\,Mpc and with redshifts differing by less than 1000\,km\,s$^{-1}$ from all other galaxies in the initial sample. At $z = 0.25$ the 1\,Mpc exclusion radius corresponds to approximately four arcminutes, which is well resolved by the ACT beam. Additional SDSS photometric data from \citet{Cunha2009} were used to account for incompleteness of nearby bright galaxies in SDSS DR7 due to fiber collisions. Stellar masses and other galaxy properties were provided by the NYU-VAGC \citep[for more details see][]{Blanton2005,Planck2013}. After all selection criteria were applied, the LBG sample used in \citet{Planck2013} contained 259,579 galaxies.

From the full LBG sample we retain only galaxies that overlap with the ACT DR6 component-separated maps and fall within the stellar mass range $\log_{10}(M_\star/M_\odot) = 11.1$ to $12.0$ reported in the NYU-VAGC. These conditions reduce the LBG sample used in this analysis to 62,677 galaxies. These LBGs are then binned by stellar mass in bins of width $\Delta\log_{10}(M_\star/M_\odot) = 0.1$ within this mass range, following the binning of \citet{Planck2013}, though starting at a slightly higher minimum stellar mass. The two most massive bins contain only 40 and 131 galaxies respectively. The estimators used in this work (described in Section~\ref{sec:meth}) assume isotropy, which is unlikely to hold for samples this small, and the sample variance in these bins is correspondingly large. We therefore drop these bins from our analysis, an issue that is compounded when we further subdivide the LBG sample by radio source presence. 

We further subdivide the LBG sample into galaxies with and without co-spatial radio sources, motivated by the residual radio source contamination identified in the component-separated Compton-$y$ maps (see Section~\ref{sec:results}). Radio sources are identified by cross-matching the LBG catalog against the NRAO VLA Sky Survey at 1.4\,GHz \citep[NVSS;][]{NVSS} using a search radius of one arcminute. The total fraction of LBGs with a co-spatial radio source is 17\%, with the remainder forming the radio source cut subsample used throughout this analysis. We note that the fraction of LBGs with a co-spatial radio source increases in the higher stellar mass bins, to around 30\%.

\section{Methods}
\label{sec:meth}

The Compton-$y$ signal we are trying to measure has low signal-to-noise from individual galaxies. We therefore use cross-correlations (colloquially referred to as stacking) to statistically extract the average signal from ensembles of galaxies, which yields higher signal-to-noise than individual detections. For every galaxy in the sample (or subsample) we use its coordinates to excise a $30' \times 30'$ sub-map from the ACT maps (component-separated and raw frequency maps) centered on that position. These sub-maps are saved where applicable for uncertainty estimation via bootstrap resampling. The sub-maps are then co-added with equal weight and without reorientation, so that for a sufficient number of galaxies the co-added sub-map is effectively isotropized. Examples of co-added sub-maps for various component-separated ACT maps are shown in Figure~\ref{fig:gallery}.

Each row of Figure~\ref{fig:gallery} shows co-adds for a particular component-separated map, as described in Section~\ref{sec:ILC}. From top to bottom the rows show: the fiducial ACT ILC (row 1), the CIB-deprojected ILC (rows 2, 4, and 5), and the moment-based CIB-deprojected ILC (row 3). Rows 4 and 5 further subdivide the LBG sample into co-adds excluding galaxies with co-spatial radio sources and co-adds including only galaxies with co-spatial radio sources, respectively. The columns subdivide the LBG sample into stellar mass bins as described in Section~\ref{sec:LBG}, with the leftmost column spanning $11.1 < \log_{10}(M_\star/M_\odot) < 11.2$ and the rightmost spanning $11.7 < \log_{10}(M_\star/M_\odot) < 11.8$, increasing in steps of 0.1\,dex from left to right. This binning follows \citet{Planck2013}.  The stellar mass bins with $\log_{10}(M_\star/M_\odot) < 11.1$ are noise dominated and we do not show them.

From the co-added maps, we extract fluxes using a compensated aperture photometry (CAP) filter following \citet{Schaan2021}. For a given aperture radius $\theta_{\text{ap}}$ centered on the co-added map, the CAP filter is defined as:
\begin{equation}
\label{eq:filt}
X(\theta_{\text{ap}}) = \int \mathrm{d}^2\theta\, X(\theta)\, W_{\theta_{\text{ap}}}(\theta),   
\end{equation}
\noindent where $W_{\theta_{\text{ap}}}$ is
\begin{equation}
    W_{\theta_{\text{ap}}} = \begin{cases} 
1 & \text{for } \theta < \theta_{\text{ap}}, \\
-1 & \text{for } \theta_{\text{ap}} \leq \theta \leq \sqrt{2}\,\theta_{\text{ap}},  \\
0 & \text{otherwise}.
\end{cases}
\end{equation}
This filter measures the map quantity $X$ within a disk of radius $\theta_{\text{ap}}$ and subtracts a concentric annulus of equal area between $\theta_{\text{ap}}$ and $\sqrt{2}\,\theta_{\text{ap}}$. The extracted fluxes $X$ are the dimensionless Compton-$y$ and temperature for the component-separated $y$-maps and raw ACT frequency maps, respectively, giving units of arcmin$^2$ and $\mu$K\,arcmin$^2$ after applying Eq.~\ref{eq:filt}.

Since we use the CAP filter for flux extraction, we do not have to assume any Compton-$y$ profile, allowing direct comparison between different ILC methods and subsamples. The aperture radius of three arcminutes is chosen to balance capturing the majority of the signal from the source while minimizing background contamination. The beams of the component-separated maps are symmetric by construction \citep{Coulton24}, so no asymmetry correction is required, and the signals are not sufficiently extended to warrant further corrections. Uncertainties on the extracted fluxes are computed using bootstrap resampling, where the individual CAP fluxes for a given stellar mass bin are randomly resampled with replacement. This method captures both the noise in the component-separated maps and the sample variance, the latter of which dominates in each stellar mass bin due to the relatively small number of galaxies.

In the previous Planck analysis of the LBG sample \citep{Planck2013}, the primary results were obtained using a multi-frequency matched filter (MMF), an optimal linear filter designed to extract the tSZ signal from multi-frequency maps given an assumed pressure profile and a noise model \citep{Melin2006}. The filter simultaneously exploits the known frequency dependence of the tSZ effect and the spatial profile of the signal to maximize the signal-to-noise of the extraction, with combination weights set by the tSZ frequency spectrum. A key assumption of the MMF is that the pressure profile of the halo is known; in the Planck analysis the so-called universal pressure profile (UPP) was assumed \citep{UPP}. Since the Planck beam is large compared to the projected virial radii of the LBGs, the MMF extracted tSZ fluxes at $5R_{500}$ and then applied a geometric correction assuming the UPP to convert these to fluxes at $R_{500}$. A follow-up analysis by \citet{LeBrun2014} using simulations showed that the assumptions underlying this correction can introduce biases in the inferred tSZ fluxes at $R_{500}$.

In contrast, our CAP-based analysis avoids these assumptions, though at the cost of optimality. This approach was also used in \citet{Greco2015} and as an additional check in the original Planck analysis \citep{Planck2013}. To make a direct comparison with \citet{Planck2013}, we alter the original Planck measurements by first undoing the MMF corrections, reverting to the tSZ fluxes at $5R_{500}$, and applying the corrections from \citet{LeBrun2014}. This effectively normalizes the pressure profile for each LBG, after which we assume the UPP profile and perform aperture photometry on the resulting profiles. Unless stated otherwise, all Planck results shown in this paper have had these corrections applied.

\section{Results}
\label{sec:results}

Figure~\ref{fig:gallery} shows a gallery of co-added sub-maps following the method described in Section~\ref{sec:meth}, with galaxies divided into stellar mass bins of width $0.1$ in $\log_{10}(M_\star/M_\odot)$. The layout of Figure~\ref{fig:gallery} is described in detail in Section~\ref{sec:meth}, with columns showing increasing stellar mass from left to right and rows showing different Compton-$y$ maps or radio source subselections. Before applying the CAP filter to extract fluxes from the co-added maps, there are already features and trends in Figure~\ref{fig:gallery} that are worth noting.

It is clear that the moment-based CIB-deprojected co-added maps have higher noise levels than the other maps, as evidenced by the larger noise fluctuation amplitudes in the co-added sub-maps. This is expected, since the additional degrees of freedom used by the moment deprojection method increase the noise in the reconstruction \citep{Chluba2017}. Despite this, the moment deprojection method has the advantage of being more robust to residual CIB contamination, since the CIB SED is not known a priori and the results may otherwise be sensitive to the choice of CIB model parameters \citep[e.g.,][]{McCarthyHill2024}. We test the robustness of our measurements to residual CIB contamination and the choice of CIB model parameters in Section~\ref{sec:fgtest}.

Another prominent feature in Figure~\ref{fig:gallery} is a negative source at the center of the co-adds, which persists even in the moment-deprojected maps, suggesting that it is not associated with dust emission. This prompted us to further divide the LBG sample into galaxies with and without co-spatial radio sources. Radio point sources co-spatial with galaxies in the cross-correlation will coherently combine their signal across the stack, appearing as a negative decrement in the Compton-$y$ maps (see Appendix~\ref{sec:sys}). These sources would be low-frequency point sources not detected in the ACT frequency maps at 90-220 GHz. In Figure~\ref{fig:gallery} the negative source is absent in the subsample excluding galaxies with co-spatial radio sources (RS Cut; Row 4), but clearly present in the subsample of galaxies with co-spatial radio sources (RS; Row 2). Radio source contamination is therefore clearly present in the ACT Compton-$y$ maps and was likely present at some level in previous analyses using these maps. As we argue in Appendix~\ref{sec:sys}, this contamination is consistent with synchrotron emission rather than an atypical dust component associated with the radio sources.

\begin{figure*}[ht!]
\centering      
    \includegraphics[width=0.98\textwidth]{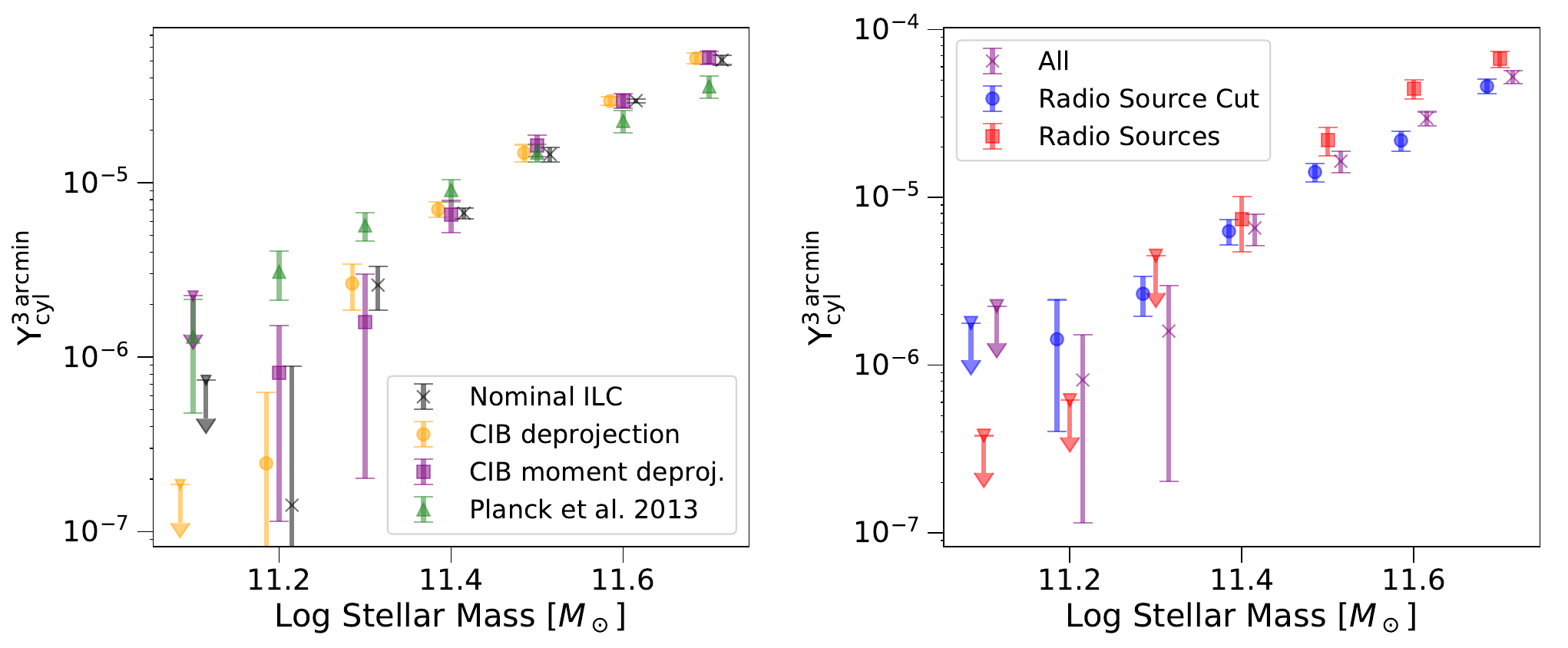}
\caption{Cylindrically integrated Compton-$y$ signal within 3 arcmin ($Y^{\mathrm{3\,arcmin}}_{\mathrm{cyl}}$) as a function of stellar mass for LBGs. \textit{Left}: Comparison of Compton-$y$ measurements from different component-separated $y$-maps for the full LBG sample: the nominal ILC (black crosses), CIB-deprojected ILC (orange circles), and moment-based CIB deprojection (purple squares), alongside the modified Planck MMF results from \citet{Planck2013} (green triangles). The three ACT-based measurements are consistent with one another across all stellar mass bins, while the Planck MMF results show comparable fluxes but with a systematically different slope as a function of stellar mass.\textit{Right}: Compton-$y$ signal from the moment-based CIB-deprojected ILC maps for the full LBG sample (purple crosses), the radio-source-cut subsample (blue circles), and the radio-source-included subsample (red squares). Galaxies hosting co-spatial radio sources show a systematically higher $Y$ signal than those without, across all stellar mass bins where a significant detection is made. Error bars denote $1\sigma$ uncertainties and downward-pointing arrows indicate $3\sigma$ upper limits.}
\label{fig:ycyl}
\end{figure*}

Having identified these features in Figure~\ref{fig:gallery}, we now quantify the Compton-$y$ signal as a function of stellar mass using the CAP filter and compare with the previous Planck results on the LBG sample. Recall that the \cite{Planck2013} tSZ measurement on the LBG sample showed a remarkably tight power-law behavior between the stellar mass of the galaxy and the tSZ signal over a decade in mass range. This result was independently confirmed by \citet{Greco2015}. Figure~\ref{fig:ycyl} shows the cylindrical Compton-$y$ signal within three arcmin as a function of stellar mass for the full LBG sample and the two subsamples of galaxies with co-spatial radio sources and those without. The left panel compares the Compton-$y$ measurements from the three ACT-based $y$-maps (the nominal ILC, the CIB-deprojected ILC, and the moment-based CIB deprojection), alongside the modified Planck MMF results from \citet{Planck2013}. The three ACT-based measurements are consistent with one another across all stellar mass bins, demonstrating that our results are fairly robust to the choice of component separation method. For all results, unless stated otherwise we use the moment-based CIB deprojection.  The stability of the moment-based CIB deprojection is discussed in Section~\ref{sec:fgtest}. The Planck MMF results show comparable fluxes to our ACT-based measurements at high stellar masses but with a systematically different slope as a function of stellar mass, with the Planck results relatively higher at lower stellar masses and lower at higher stellar masses. The difference in slopes is likely driven by the difference in beams between ACT and Planck, which will impact our conversion from the original MMF fluxes to CAP fluxes within three arcminutes, including the corrections proposed in \citet{LeBrun2014} described in Section~\ref{sec:meth}. We still find a power-law relation, albeit over a smaller range in stellar mass, in part resulting from a smaller sample and less optimal flux extraction. 

\begin{table*}[ht!]
\centering
\begin{tabular}{c|c|c|c|c|c|c}
\hline
$\mathrm{Log}_{10}(M_\star)$ & U & $\mathrm{Y}_{\mathrm{cyl}}^{3'}$ Radio Cut  & $E_{th}$ & $\mathrm{Y}_{\mathrm{cyl}}^{3'}$ Radio  &  $E_{th}$ & $\Delta E_{th}$ \\
$[M_{\odot}]$ & [$10^{62}$ erg] & [$10^{-6}$ arcmin$^{2}$] & [$10^{62}$ erg]  & [$10^{-6}$ arcmin$^{2}$] & [$10^{62}$ erg] & [$10^{62}$ erg] \\
\hline
11.2 - 11.3 & 1.3 &1.4 $\pm$ 1.0 & 0.7 $\pm$ 0.5 & 4.4 $\pm$ 1.4 & 2.3 $\pm$ 0.7 & 1.6 $\pm$ 0.9 \\
11.3 - 11.4 & 3.3 &2.8 $\pm$ 0.7 & 1.4 $\pm$ 0.4 & 6.1 $\pm$ 2.7 & 3.2 $\pm$ 1.4 & 1.8 $\pm$ 1.5 \\
11.4 - 11.5 & 8.2 &6.3 $\pm$ 1.1 & 3.2 $\pm$ 0.6 & 17.4 $\pm$ 3.0 & 8.9 $\pm$ 1.6 & 5.7 $\pm$ 1.7 \\
11.5 - 11.6 & 18 &14.0 $\pm$ 1.7 & 7.3 $\pm$ 0.9  & 28.9 $\pm$ 4.7 & 14.9 $\pm$ 2.4 & 7.6 $\pm$ 2.6 \\
11.6 - 11.7 & 40 &21.8 $\pm$ 3.0 & 11.2 $\pm$ 1.6 & 48.8 $\pm$ 6.4 & 25.1 $\pm$ 3.3 & 13.9 $\pm$ 3.6 \\
11.7 - 11.8 & 83 &46.0 $\pm$ 4.6 & 23.6 $\pm$ 2.4 & 70.3 $\pm$ 8.5 & 36.2 $\pm$ 4.4 & 12.6 $\pm$ 5.0 \\
\hline
\end{tabular}
\caption{Cylindrical Compton-$y$ signal ($Y_{\mathrm{cyl}}^{3'}$) and corresponding thermal energy ($E_{th}$) measured within a three arcminute aperture from the moment-based CIB-deprojected ACT $y$-map ($\partial\beta$, $\beta=1.6$, $T^{\mathrm{eff}}_{\mathrm{CIB}}=24$\,K) for LBGs without co-spatial radio sources (Radio Cut) and with co-spatial radio sources (Radio), in bins of stellar mass. The second column contains the estimate of the binding energy, $U$. The $Y_{\mathrm{cyl}}^{3'}$ and $E_{th}$ values for the radio source subsample (labeled Radio) have been corrected for central point source contamination by subtracting the difference in CAP fluxes within $2'$ between the two subsamples. The final column gives the difference in thermal energy, $\Delta E_{th}$, between the corrected radio source subsample and the radio source cut subsample. We note that $E_{th}$ is computed from the corrected $Y_{\mathrm{cyl}}^{3'}$ via Equation~\ref{eq:eth} and all uncertainties are $1\sigma$.}\label{tab:meas}
\end{table*}

The right panel of Figure~\ref{fig:ycyl} shows the moment-based CIB-deprojected Compton-$y$ signal for the full LBG sample and the two subsamples defined by the presence or absence of co-spatial radio sources. Across all stellar mass bins where a significant detection is made, galaxies hosting co-spatial radio sources show a systematically higher Compton-$y$ signal than those without. At the highest stellar mass bins, $\log_{10}(M_\star/M_\odot) \gtrsim 11.4$, both subsamples are detected at high significance and the difference between them is clearly resolved. At lower stellar masses, the signal-to-noise decreases and the measurements for the radio source subsample are consistent with upper limits, while the radio source cut subsample remains detected. The full sample sits between the two subsamples as expected, reflecting the weighted contribution of each to the average of the full sample.

These results show that galaxies within a fixed stellar mass bin that have co-spatial radio sources contain more Compton-$y$ signal than galaxies without radio sources. This is quantified in Table~\ref{tab:meas}, which lists for each $M_\star$ bin with a measurement the average $\mathrm{Y}_{\mathrm{cyl}}^{3'}$ without and with co-spatial radio sources for the moment-based CIB-deprojected Compton-$y$ map. In Figure~\ref{fig:ycyl} we note that there are no corrections to account for the missing Compton-$y$ signal in the subsample with co-spatial radio sources due to the residual contamination filling in the tSZ decrement at frequencies below 220 GHz. This is not the case in Table~\ref{tab:meas}, where we use the flux from two arcminutes in the radio-source-cut subsample as the {\it true} Compton-$y$ signal to estimate the correction due to radio emission contamination in the other sample. We emphasize that this simple correction, which takes the difference in fluxes within two arcminutes as the amount of {\it missing} Compton-$y$ flux, is an estimate and could be subject to systematic offsets of around 50\%. With these corrections we find that within three arcmin there is roughly a factor of two more $\mathrm{Y}_{\mathrm{cyl}}^{3'}$ from LBGs with co-spatial radio sources compared to those without these sources. Regardless, without these corrections, sources with co-spatial radio sources still show more Compton-$y$ flux than those without. In Appendix~\ref{sec:sys}, we test that the excess Compton-$y$ signal ($\mathrm{Y}_{\mathrm{cyl}}^{3'}$) in LBGs with co-spatial radio sources is intrinsic to those galaxies and discuss the possible reasons for these differences in Section~\ref{sec:disc}.

\begin{figure}[t!]
\centering          \includegraphics[width=0.49\textwidth]{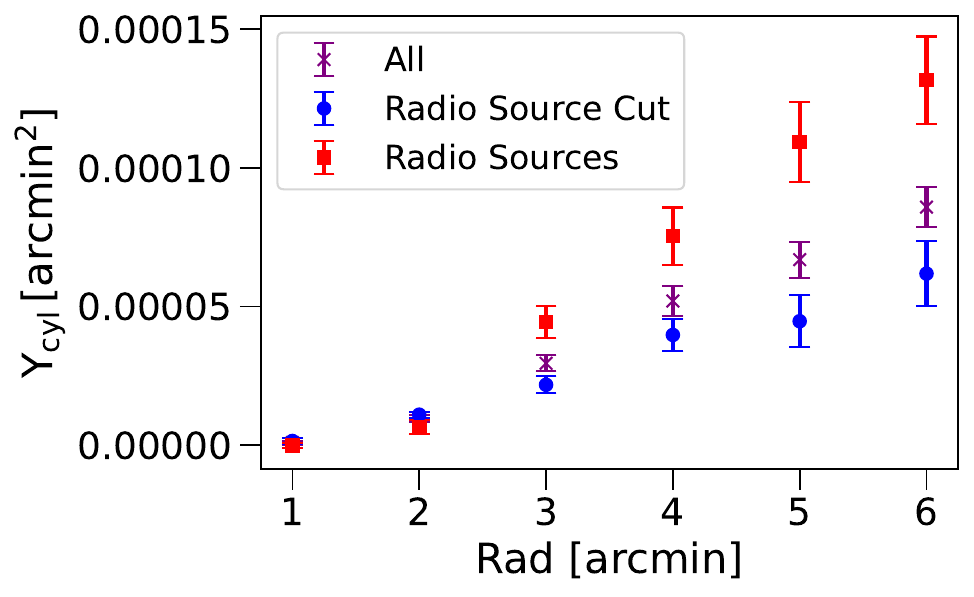}
\caption{Radial profile of the cylindrical Compton-$y$ signal, $Y_{\mathrm{cyl}}$ 
(arcmin$^2$), as a function of aperture radius for the full LBG sample (purple crosses), the radio-source-cut subsample (blue circles), and the radio source subsample (red squares), for the stellar mass bin $11.5 < \log_{10}(M_\star/M_\odot) < 11.6$, measured from the moment-based CIB-deprojected ACT $y$-map. The radio source subsample shows systematically higher $Y_{\mathrm{cyl}}$ than the radio-source-cut subsample at all radii beyond two arcminutes, with the excess persisting out to the maximum radius considered in this analysis of six arcminutes. Error bars denote $1\sigma$ uncertainties.}
\label{fig:radprof}
\end{figure}

The choice of a three arcminute aperture for our primary CAP measurements is not physically special, and Figure~\ref{fig:radprof} shows that the excess Compton-$y$ signal in galaxies with co-spatial radio sources is not confined to this scale. Figure~\ref{fig:radprof} shows the radial profile of $Y_{\mathrm{cyl}}$ as a function of aperture radius for the full sample and both subsamples in the stellar mass bin $11.5 < \log_{10}(M_\star/M_\odot) < 11.6$. At radii of $1$--$2$ arcminutes the sample with radio sources is lower than the other two, reflecting the fact that the central radio source contamination is confined to the inner aperture. Beyond two arcminutes, where the radio source contamination is no longer dominant, the radio source subsample shows a systematically higher $Y_{\mathrm{cyl}}$ than the radio-source-cut subsample, and this excess persists out to six arcminutes, the maximum radius considered in this analysis. The full sample sits between the two subsamples at all radii, as expected.

The persistence of this excess to large radii suggests that the difference between the two subsamples is not simply a central point source effect but reflects a genuine difference in the Compton-$y$ signal on halo scales. We note that no flux corrections have been applied to these profiles. As we show in Appendix~\ref{sec:sys}, this excess is not attributable to residual ILC contamination, further supporting the interpretation that the excess Compton-$y$ in galaxies hosting co-spatial radio sources is intrinsic to those galaxies.

\subsection{Robustness to CIB Contamination}
\label{sec:fgtest}

\begin{figure*}[ht!]
\centering      
    \includegraphics[width=0.98\textwidth]{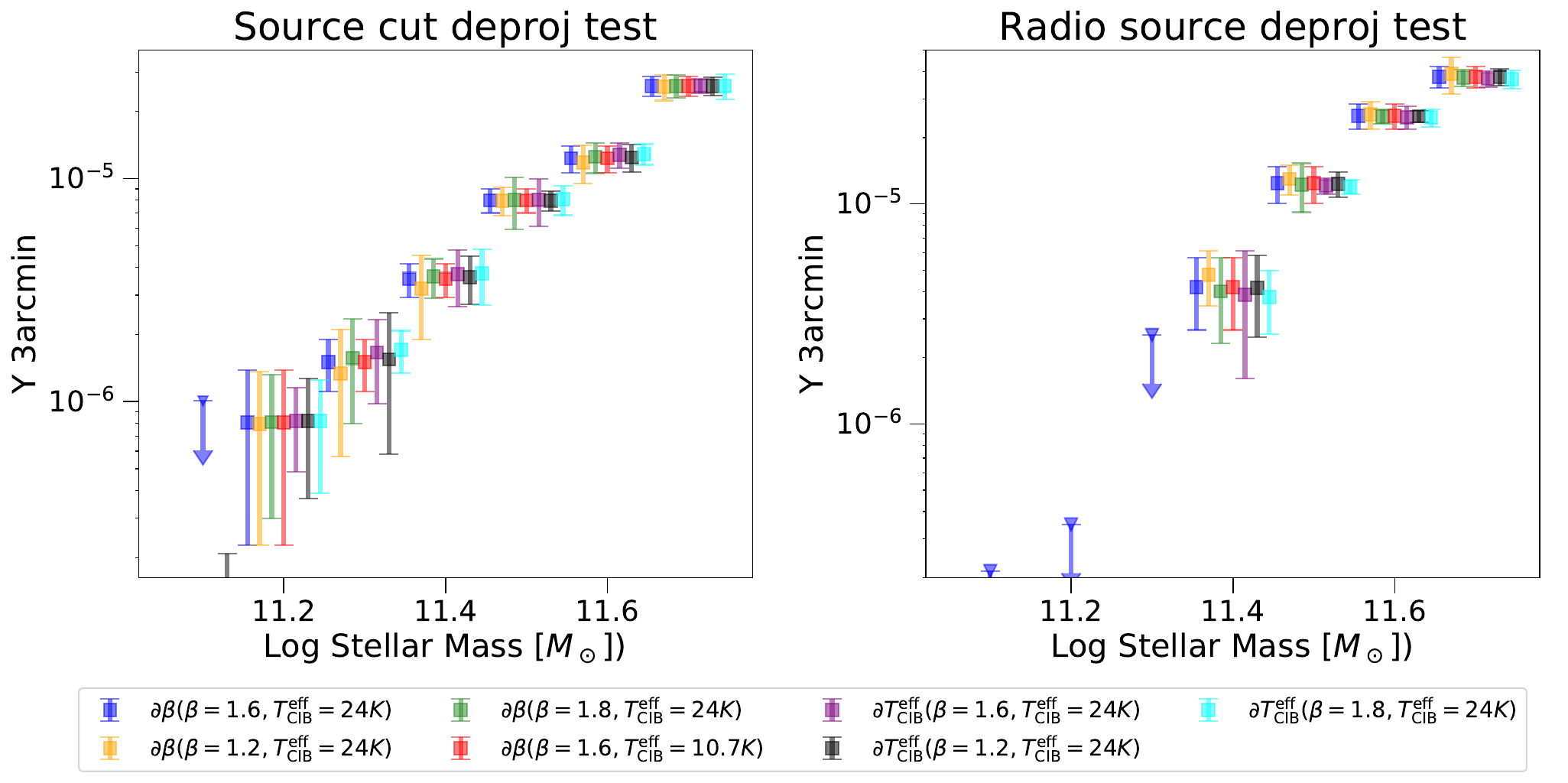}
\caption{Cylindrical Compton-$y$ signal ($Y^{\mathrm{3\,arcmin}}_{\mathrm{cyl}}$) as a function of stellar mass for the radio-source-cut subsample (\textit{left}) and the radio source subsample (\textit{right}) of LBGs, measured using the moment-based CIB-deprojected $y$-maps with varying CIB SED parameters. Results are shown for first-moment deprojections in $\beta$ ($\partial\beta$) and $T^{\mathrm{eff}}_{\mathrm{CIB}}$ ($\partial T^{\mathrm{eff}}_{\mathrm{CIB}}$), across spectral indices $\beta \in \{1.2, 1.6, 1.8\}$ and effective temperatures $T^{\mathrm{eff}}_{\mathrm{CIB}} \in \{10.7, 24\}$\,K. The consistency of the measured Compton-$y$ signal across all parameter choices in both subsamples demonstrates convergence of our results with respect to the assumed CIB SED, and confirms that the difference between the two subsamples is not driven by the choice of deprojection parameters.}
\label{fig:fgtest}
\end{figure*}

A concern for all measurements on component-separated tSZ maps is the sensitivity of the result to residual CIB contamination in these maps. The moment-based CIB deprojection method aims to mitigate this by suppressing CIB residuals without requiring precise knowledge of the CIB SED parameters a priori. We verify that our results have converged by checking that the measured Compton-$y$ signal is stable under different choices of CIB SED parameters used to define the moments, namely, the spectral index $\beta$ and effective CIB temperature $T^{\mathrm{eff}}_{\mathrm{CIB}}$. To test this, we repeat our Compton-$y$ measurements using a range of CIB SED parameters, varying $\beta \in \{1.2, 1.6, 1.8\}$ and $T^{\mathrm{eff}}_{\mathrm{CIB}} \in \{10.7, 24\}$\,K, and deprojecting both the first moment in $\beta$ ($\partial\beta$) and the first moment in $T^{\mathrm{eff}}_{\mathrm{CIB}}$ ($\partial T^{\mathrm{eff}}_{\mathrm{CIB}}$). Figure~\ref{fig:fgtest} shows the results of this test for both the co-spatial radio-source-cut subsample (left) and the radio source subsample (right). In both cases the measured Compton-$y$ signal is consistent across all choices of CIB parameters, demonstrating that our results have converged with respect to the assumed CIB SED parameters. This stability under different deprojection parameter choices strengthens our confidence that residual CIB is well within our statistical uncertainties and that our measured Compton-$y$ signals are not biased by these residuals.

\section{Discussion}

\label{sec:disc}

A central result of this work is that LBGs hosting co-spatial radio sources show a systematically higher Compton-$y$ signal than those without, with roughly a factor of two difference in $Y_{\mathrm{cyl}}^{3'}$ at fixed stellar mass across multiple stellar mass bins. We establish through several tests in Appendix~\ref{sec:sys} that the excess Compton-$y$ signal is intrinsic to LBGs with co-spatial radio sources. We also note that the excess Compton-$y$ in LBGs with co-spatial radio sources is not confined to the central aperture but persists to radii of at least six arcminutes (Figure~\ref{fig:radprof}), the maximum radius considered in this analysis. We consider two physical explanations for this excess: that LBGs with co-spatial radio sources reside in systematically more massive halos, or that the excess reflects energy injected into the CGM by AGN feedback. Both explanations are consistent with the excess Compton-$y$ signal persisting to large radii.

One possible explanation for this excess is that LBGs with co-spatial radio sources are systematically more massive than those without. Assuming $\mathrm{Y}_{\mathrm{cyl}}^{3'}$ scales with halo mass similarly to the self-similar scaling of the spherically integrated $Y$ signal~\citep[e.g.,][]{UPP}, LBGs with co-spatial radio sources would need to be $\sim$50\% more massive to account for a factor of two difference in $\mathrm{Y}_{\mathrm{cyl}}^{3'}$. In Appendix~\ref{sec:sys}, we show that the stellar mass distributions of LBGs with and without co-spatial radio sources are statistically indistinguishable. For a higher halo mass to explain the observed difference in $\mathrm{Y}_{\mathrm{cyl}}^{3'}$, LBGs with co-spatial radio sources would therefore need to be systematically $\sim$50\% more massive in halo mass while maintaining the same stellar mass as their counterparts without radio sources.

A previous galaxy-galaxy lensing and clustering analysis of SDSS galaxies by \citet{Mandelbaum2008} showed that at fixed stellar mass, galaxies hosting radio-loud AGN live in more massive halos than those without. This would imply that the excess Compton-$y$ signal that we observe is a consequence of LBGs with co-spatial radio sources residing in more massive halos, providing additional independent evidence for the \citet{Mandelbaum2008} result. However, a subsequent galaxy clustering analysis by \citet{Worpel2013} did not reach the same conclusion. In Appendix~\ref{sec:sys}, we show that our CMB halo lensing measurement cannot statistically distinguish between the halo masses of the two subsamples. It is therefore plausible that we are preferentially selecting $\sim$50\% more massive halos when we make our radio source subselection, but we cannot conclusively establish this claim with the current data.

Another explanation for the excess is that we are observing the effects of AGN feedback on the CGM. If these samples are all at comparable halo masses then we can ask whether the energetics of AGN feedback can explain this measurement. Specifically, is there sufficient energy output from an AGN to account for the additional Compton-$y$ signal, and would this amount of thermal energy be sufficient to unbind the gas? We determine the amount of thermal energy following \citet{Ruan2015}. From Equation~\ref{eq:y} we can calculate the total Compton-$y$ in a cylinder,
\begin{equation}
    Y(z) = D_A^2 (z) \int y \,\mathrm{d}\Omega,
\end{equation}
where the integral is over solid angle $\Omega$ on the sky, $D_A(z)$ is the angular diameter distance, and $Y(z)$ carries units of Mpc$^2$.

The inferred total thermal energy ($E_{th}$) from $Y(z)$ is 
\begin{equation}
    \label{eq:eth}
    E_{th} = \frac{3 m_e c^2}{2\sigma_T}f_e Y(z),
\end{equation}
where $f_e$ is the inverse of the electron fraction ($x_e = 1.158$), defined as the ratio of electron and hydrogen number densities, times the primordial abundance of hydrogen ($X_H = 0.76$), times the mean molecular weight ($\mu = 0.588$), for a fully ionized medium of primordial abundance \citep{BBPS1}. For this calculation we use an aperture of three arcminutes and adopt $z = 0.25$, which corresponds to the peak of the redshift distribution for the stellar masses considered here. Changing $z$ scales the inferred $E_{th}$ by $D_A^2(z) / D_A^2(z = 0.25)$. In Table~\ref{tab:meas} we show for each $M_\star$ bin the average $\mathrm{Y}_{\mathrm{cyl}}^{3'}$ and $E_{th}$ for both subsamples, along with the additional $E_{th}$ for LBGs with co-spatial radio sources relative to those without. The additional $E_{th}$ is comparable in magnitude to the $E_{th}$ measured from LBGs without co-spatial radio sources.

Is it possible that the radio sources in these LBGs can account for this excess energy in the CGM? The radio sources are most likely powered by supermassive black holes inefficiently accreting material and injecting energy into the surrounding CGM. There is an extensive literature on such objects and how inefficient accretion powers radio jets \citep{Narayan1994}; such systems are referred to as Advection-Dominated Accretion Flows (ADAFs), and we adopt that term here. It is commonly argued that quasars accrete near the Eddington limit, whereas ADAFs do not, which we account for with an accretion efficiency factor ($\epsilon_\mathrm{acc}$). The energy associated with such accretion can be written as some fraction of the Eddington luminosity over some timescale,
\begin{eqnarray*}
    E_\mathrm{inj} &=& \epsilon_\mathrm{acc} \dot{M}_\mathrm{Edd} c^2 \Delta t \\%
     &=& 1.5 \times 10^{62} \mathrm{erg} \left ( \frac{\epsilon_\mathrm{acc}}{0.02} 
     \right ) \left ( \frac{M_\mathrm{BH}}{10^9 M_\odot} \right ) \left ( 
     \frac{\Delta t}{2\,\mathrm{Gyr}} \right ).
\end{eqnarray*}
For our fiducial assumptions of $\epsilon_\mathrm{acc}$, $M_\mathrm{BH}$, and $\Delta t$, the injected thermal energy is $1.5 \times 10^{62}$\,erg, which is sufficient to account for the additional $E_{th}$ in most stellar mass bins. Deviations from this estimate can be reconciled by adjusting $\epsilon_\mathrm{acc}$, $M_\mathrm{BH}$, or $\Delta t$, each of which scales linearly with $E_\mathrm{inj}$. In these calculations we assume that ADAF radio sources in these galaxies do not have a duty cycle analogous to quasars, i.e., once they turn on, they remain on.

The final energetics question is whether the hypothesized energy injected into the CGM is capable of unbinding this gas. We estimate the average binding energy of the LBGs from the stellar-to-halo mass relation computed in \citet{Planck2013}, which uses the Millennium Simulation \citep{Springel2005} rescaled to the WMAP7 cosmology \citep{Angulo2010} and the semi-analytic galaxy formation model from \citet{Guo2011}. The binding energy of a halo is
\begin{equation}
    U \simeq \frac{GM^2}{R},
\end{equation}
where $G$ is Newton's gravitational constant and $M$ and $R$ are the mean halo mass and radius of LBGs in a given stellar mass bin, respectively. There is an order-unity prefactor that we neglect. Table~\ref{tab:meas} shows that in all mass bins the binding energies are comparable to or more than the total thermal energy in these systems, suggesting the injected energy is insufficient to unbind the gas. We note that the halo masses are calculated from the stellar-to-halo mass relation derived in \citet{Planck2013}. This calculation carries uncertainties of roughly a factor of two, particularly given that the two subsamples may not have the same halo masses at fixed stellar mass.

We have chosen not to compare our results to state-of-the-art cosmological simulations of galaxy formation \citep[e.g.,][]{TNG,Simba,Flamingos}, as has been done for previous SZ measurements \citep[e.g.,][]{Amodeo2021,Moser2022,Hadzhiyska2025,Siegel2026}. While such comparisons are valuable and test the sub-grid models within these simulations, it is critical that the galaxy sample is modeled correctly \citep{Moser2021,Pandey2022,McCarthy2025,Popik2025,Kadir2026}, and faithfully reproducing the LBG selection is non-trivial. The LBG sample is defined by cuts on galaxy colors, an isolation criterion, and as we have shown here, the fraction of galaxies hosting co-spatial radio sources, which directly impacts the measured Compton-$y$ signal. Correctly modeling all of these selection effects simultaneously in a simulation is a substantial undertaking and warrants a dedicated study in its own right.

\section{Conclusion}
\label{sec:conc}

In this work we measured the thermal Sunyaev-Zel'dovich (tSZ) effect from the Locally Bright Galaxy (LBG) sample using the ACT DR6 component-separated $y$-maps \citep{Coulton24}, revisiting the original Planck analysis of \citet{Planck2013} with significantly improved angular resolution. We find a clear power-law scaling between the cylindrical Compton-$y$ signal and stellar mass, consistent with the original Planck result, albeit over a smaller stellar mass range due to the smaller sky coverage of ACT relative to Planck, and our use of a less optimal but assumption-free flux extraction method. The three ACT-based measurements, the nominal ILC, CIB-deprojected ILC, and moment-based CIB deprojection, are consistent with one another across all stellar mass bins, demonstrating robustness to the choice of component separation method. The conservative moment-based CIB deprojection results are presented as our final results, which sacrifices signal-to-noise for robustness to CIB residuals.

The central result of this work is the identification of a systematic excess in the Compton-$y$ signal in LBGs hosting co-spatial radio sources relative to those without, amounting to roughly a factor of two in $Y_{\mathrm{cyl}}^{3'}$ at fixed stellar mass. Through a series of systematic tests detailed in Appendix~\ref{sec:sys}, we establish that this excess is intrinsic to the galaxies and not an artifact of residual contamination in the component-separated $y$-maps.  We find this same excess when applying the radio source subselection to the original Planck MMF results. The excess persists to radii of at least $6'$, suggesting it is a halo-scale effect independent of the central negative residual introduced by point source contamination. We consider two physical explanations: that LBGs with co-spatial radio sources reside in systematically more massive halos at fixed stellar mass, consistent with the findings of \citet{Mandelbaum2008}, or that the excess reflects thermal energy injected into the CGM by AGN feedback in the form of ADAFs. A simple energetics argument shows that AGN feedback can plausibly account for the observed excess thermal energy, and that this energy is comparable to but unlikely to exceed the binding energy of the host halos. Both explanations are well motivated and consistent with the data, and we cannot currently distinguish between them.

This result has direct implications for previous tSZ measurements using ACT $y$-maps and for tSZ cross-correlation measurements more broadly. Radio source contamination of the kind identified here will be present at some level in cross-correlations of 90\,GHz, 150\,GHz, and component-separated maps. Low-frequency channels with comparable resolution would greatly improve one's ability to remove radio source contributions from tSZ flux measurements, but this would require large-aperture centimeter-wave, survey-capable telescopes that do not currently exist. A more realistic direction is to push to lower frequency radio surveys ($\sim$1\,GHz) to identify all possible sources of radio contamination, with projects such as the DSA-2000 \citep{DSA2000} coming online in the next several years.

In addition to contamination, radio sources within galaxies pose questions about how to proceed with theoretical modeling and comparisons to simulations. We clearly show that at fixed stellar mass, galaxy samples with co-spatial radio sources have a higher measured Compton-$y$ signal than those without. This effect is particularly relevant at the high stellar mass end where radio-loud AGN are more commonly observed. How this should be accounted for in future tSZ analyses will depend on the question being asked. What is clear is that careful characterization of the radio source fraction in galaxy samples used for tSZ cross-correlations is warranted, and that radio source subselection should be considered a standard systematic check in such analyses. Going forward, modeling the LBG selection, including the galaxy color cuts, isolation criterion, and radio source fraction, in state-of-the-art cosmological simulations would allow a more direct comparison with theoretical predictions.

Residual dust contamination remains a major concern for tSZ measurements on galaxy samples. Moment-based ILC methods mitigate this by suppressing CIB residuals without requiring precise knowledge of the dust SED, but at the cost of a significant noise penalty. Access to higher frequency channels would help on both fronts, improving dust mitigation while also reducing the noise penalty associated with moment deprojection. The Fred Young Submillimeter Telescope (FYST), constructed by the CCAT collaboration, will provide higher frequency coverage that will directly address this \citep{CCAT}, and the Simons Observatory will add a 280\,GHz channel that will similarly improve dust characterization in future analyses \citep{SO2019,SO2025}.

An interesting open question from this work is whether the excess Compton-$y$ signal in LBGs with co-spatial radio sources is driven by a difference in halo mass or by AGN feedback. Galaxy-galaxy lensing measurements of the two subsamples would provide a direct constraint on their halo masses at fixed stellar mass, and potentially allow us to distinguish between these two explanations. Current surveys such as the Dark Energy Survey \citep[DES;][]{DES} and the Kilo-Degree Survey \citep[KiDS;][]{KiDS} will provide higher signal-to-noise than the CMB lensing measurement shown in Appendix~\ref{sec:sys}. Once the Rubin Observatory Legacy Survey of Space and Time \citep[LSST;][]{LSST} and the \textit{Euclid} space telescope \citep{Euclid} begin publishing their large-area deep imaging data, they will provide higher signal-to-noise galaxy-galaxy lensing measurements than current surveys and further address this open question.

\begin{acknowledgments}
This research was supported in part by grant no.~NSF PHY-2309135 to the Kavli Institute for Theoretical Physics (KITP). NB acknowledges support from NASA grant 80NSSC25K7821. JCH acknowledges support from NSF grant AST-2307727, NASA grant 80NSSC23K0463 [ADAP], and the Sloan Foundation. We would like to thank James Sullivan, Mathew Madhavacheril, Simone Ferraro, Norm Murray, Jenny Greene, for their helpful comments on this work.
\end{acknowledgments}

\appendix

\section{Systematic Tests}
\label{sec:sys}

\subsection{Radio Contamination of Compton-$y$ Map}

\begin{figure*}[ht!]
\centering          \includegraphics[width=0.49\textwidth]{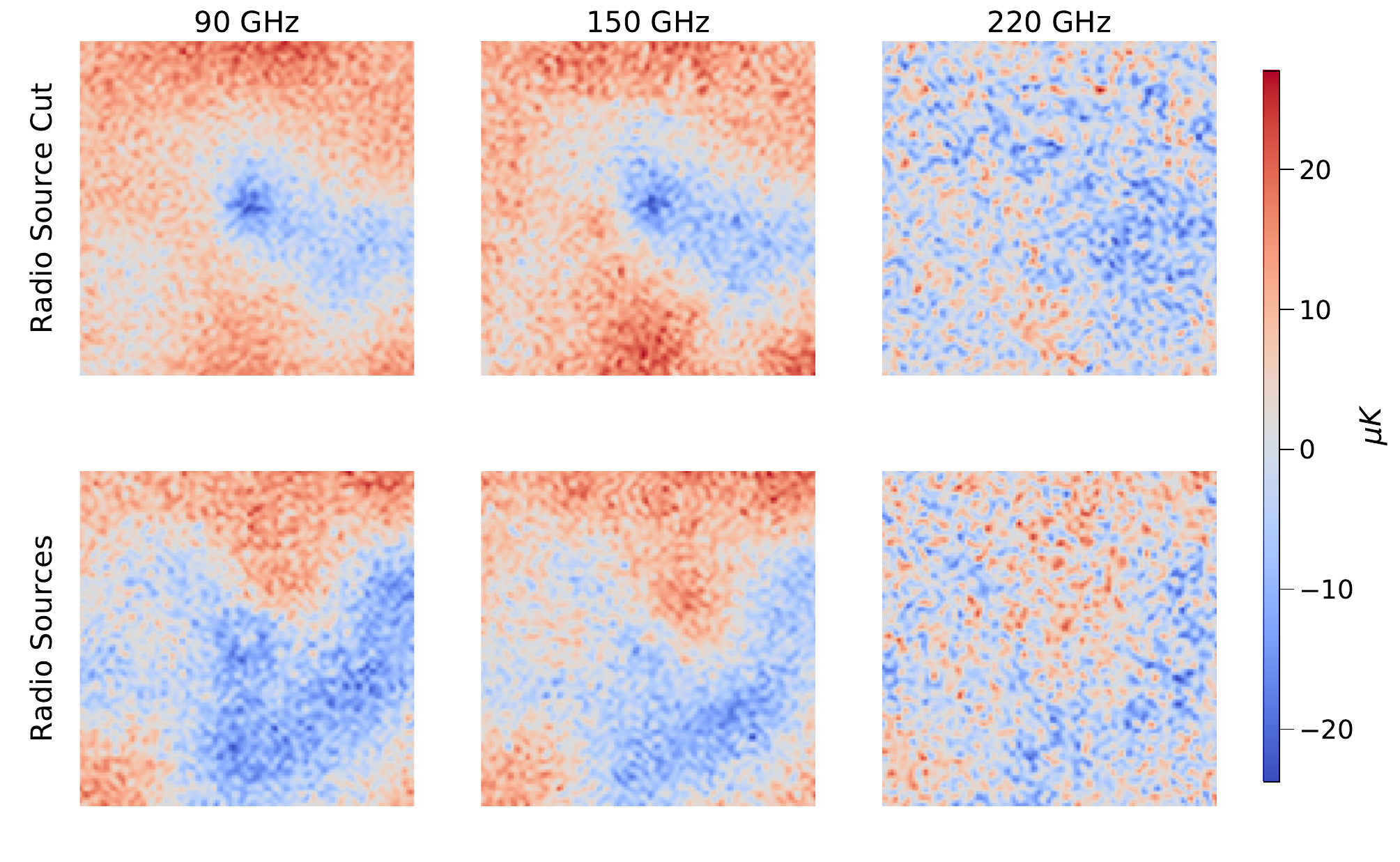}
\includegraphics[width=0.49\textwidth]{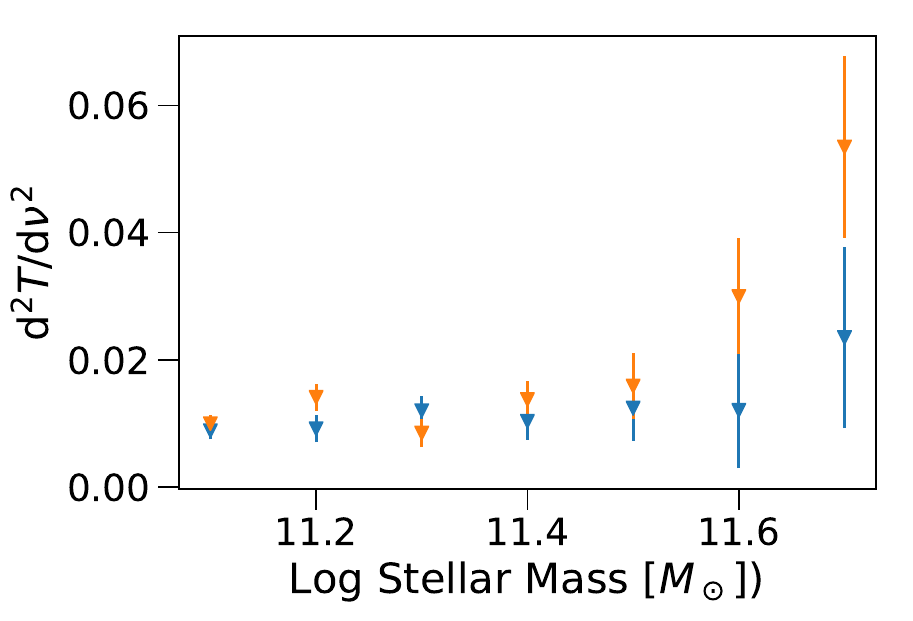}
\caption{\textit{Left}: Co-added $30 \times 30$ arcminute cutouts of single-frequency ACT maps at 90, 150, and 220\,GHz (columns) for LBGs without co-spatial radio sources (top row) and with co-spatial radio sources (bottom row), in the stellar mass bin $11.6 < \log_{10}(M_\star/M_\odot) < 11.7$. A central decrement consistent with the tSZ effect is visible in the radio-source-cut sample, while the radio source sample shows an additional central positive residual at 90 and 150\,GHz, indicative of residual radio source contamination. \textit{Right}: Numerical second derivative of the CAP-filtered temperature, $\mathrm{d}^2T/\mathrm{d}\nu^2$, as a function of stellar mass for galaxies with (orange) and without (blue) co-spatial radio sources. In all but one mass bin, the radio source subsample shows a larger second derivative, consistent with excess synchrotron emission at low frequencies from the co-spatial radio point sources.}
\label{fig:derivs}
\end{figure*}

Residual contamination is evident at the center of the stacked cutout maps for all types of component-separated $y$-maps we analyze. This is a systematic that all tSZ cross-correlation measurements must continue to mitigate, particularly as signal-to-noise ratios increase. It is clear that this residual is not present when galaxies without co-spatial radio point sources are selected. However, it is not immediately evident that the residual in the $y$-maps is caused by the radio point sources themselves. To investigate this, we perform the same CAP filter stacking analyses using single-frequency ACT maps in place of the component-separated $y$-maps. The ACT data has broadband frequencies centered at approximately 90, 150, and 220 GHz~\citep{Naess25}. The left panel of Figure~\ref{fig:derivs} shows stacked cutouts of these single-frequency maps for both subsamples, where the radio source subsample shows a central positive residual at 90 and 150\,GHz that is absent in the radio-source-cut subsample, consistent with residual synchrotron emission from the co-spatial radio point sources. We choose not to attempt a parametric fit across these single-frequency cross-correlations, as even the simplest models characterizing the tSZ, dust, and synchrotron components contain more free parameters than we have frequency bands. Instead, we numerically approximate the second derivative of the CAP-filtered temperature across the three frequency bands using finite differences, where excess emission at 90 GHz from a residual radio component would manifest as a larger second derivative, i.e., a more convex spectrum. Figure~\ref{fig:derivs} shows that in all but one mass bin, the subsample of galaxies with co-spatial radio sources has a larger second derivative than those without. This is compelling evidence that the centrally located positive residual in the $y$-maps is caused by the co-spatial radio point sources.

\subsection{Compton-$y$ Map Response to Radio Sources}

When constructing a component-separated $y$-map using an ILC method, the weights assigned to each input map determine the map's response to residual contamination from other components. In our case, the $y$-maps were constructed using a needlet ILC method, which varies its weights as a function of angular scale and spatial position. A concern is therefore that our result, showing that galaxies with co-spatial radio sources have a larger Compton-$y$ signal than those without, could be a spurious artifact of residual radio contamination leaking through the needlet ILC weights. Since the response to these spatially and frequency-dependent weights is non-trivial to calculate analytically~\citep{Surrao2024}, we devised an empirical test of whether such contamination could manifest as a spurious Compton-$y$ signal.

\begin{figure*}[t!]
\centering          
\includegraphics[width=0.49\textwidth]{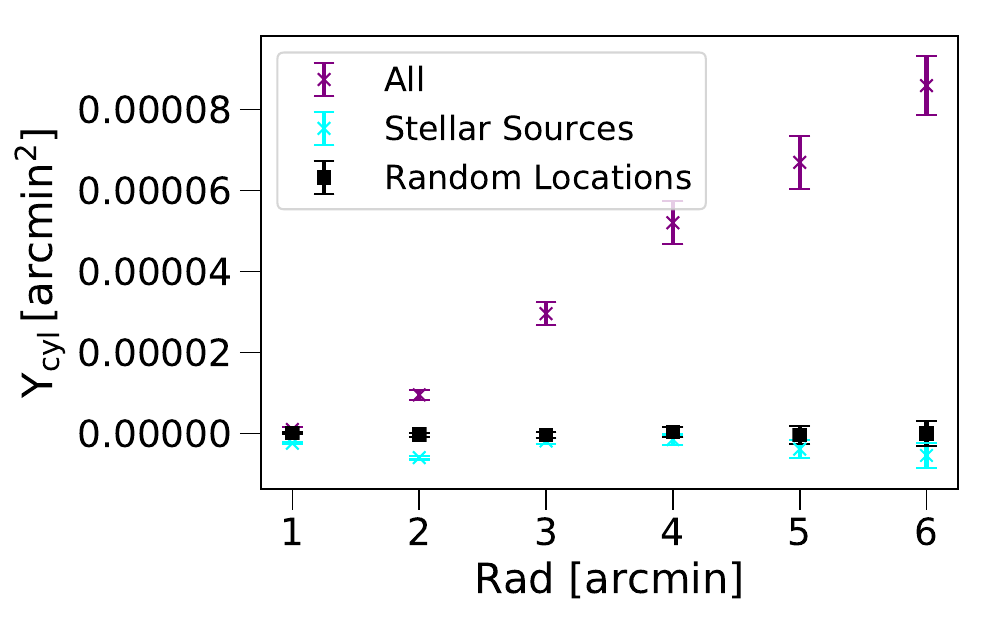}
\includegraphics[width=0.49\textwidth]{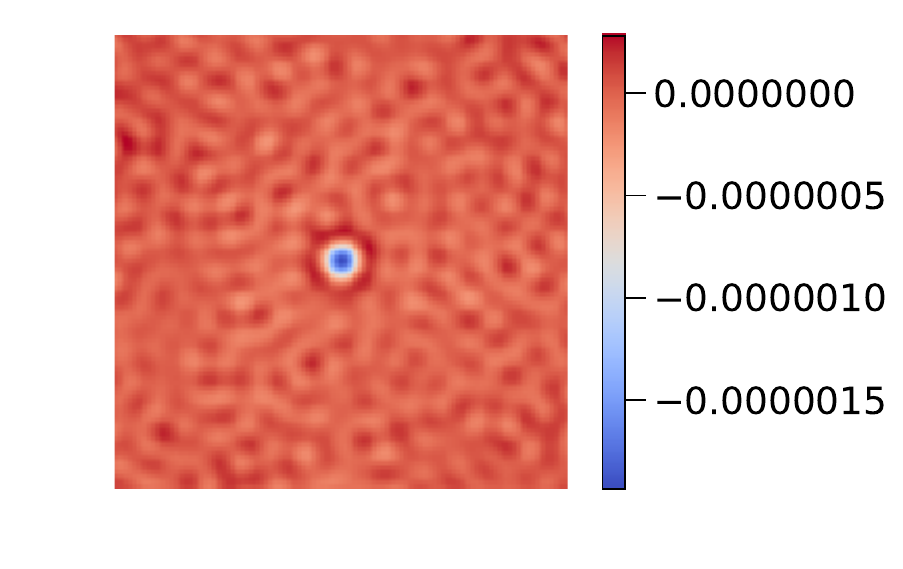}
\caption{\textit{Left}: Radial profiles of the cylindrical Compton-$y$ signal, 
$Y_{\mathrm{cyl}}$ (arcmin$^2$), as a function of aperture radius for LBGs 
in the bin $11.6 < \log_{10}(M_\star/M_\odot) < 11.7$ (purple), stellar radio sources from the FIRST catalog (cyan), and random sky locations (black). The stellar radio source profile quantifies the level of spurious Compton-$y$ from residual radio contamination in the needlet ILC $y$-map, which is more than an order of magnitude below the measured LBG signal at radii $\gtrsim 3$ arcminutes. Within one and two arcminutes the stellar radio sources have CAP fluxes of $-2.43 \pm 0.17 \times 10^{-6}$ and $-6.01 \pm 0.45 \times 10^{-6}$, respectively. \textit{Right}: $30 \times 30$ arcminute co-added cutout of the moment-CIB-deprojected $y$-map at the positions of FIRST stellar radio sources, showing the amplitude of residual radio contamination. The central decrement is consistent with a small negative residual, confirming that spurious Compton-$y$ from radio source contamination is negligible.}
\label{fig:radiotest}
\end{figure*}

We use stellar radio sources from the FIRST radio source catalog \citep{First} to cross-correlate against the $y$-maps and quantify the amplitude of any residual radio contamination. Stellar radio sources have no intrinsic Compton-$y$ signal since they do not live in massive dark matter halos, making them ideal tracers of residual radio contamination without the complication of having to subtract a true Compton-$y$ signal. Within the ACT $y$-map footprint there are over 76,900 stellar radio sources. A $30 \times 30$ arcminute cutout of the co-added on the moment CIB-deprojected $y$-map is shown in the right panel of Figure~\ref{fig:radiotest}. The positive residuals visible in this image represent spurious Compton-$y$, and their amplitude appears small by eye. We directly compare the radial profiles of the stellar radio sources, LBGs, and random locations in the left panel of Figure~\ref{fig:radiotest}. Within one and two arcminutes, the stellar radio sources have CAP fluxes of $-2.43 \pm 0.17 \times 10^{-6}$ and $-6.01 \pm 0.45 \times 10^{-6}$, respectively. We restrict the flux calculation to aperture sizes comparable to the beam because point sources are unresolved at these scales. These fluxes are predominantly greater than or equal to the residual contamination fluxes in the LBGs with co-spatial radio sources, even after the intrinsic Compton-Y signal of the LBGs has been subtracted off. Thus, our residual test using these stellar radio sources is conservative, given that the flux from stellar radio sources is larger than or equal to that from the co-spatial LBG radio sources. The random location cross-correlation, computed using the same number of points as the galaxy samples, quantifies the noise level in these cross-correlations and serves as a baseline for assessing significance. Figure~\ref{fig:radiotest} shows that the residual spurious Compton-$y$ from the stellar radio source cross-correlation is more than an order of magnitude below our measured Compton-$y$ signal at radii of $\gtrsim 3$ arcminutes, where radio source contamination would be expected to manifest. The LBGs shown in Figure~\ref{fig:radiotest} span the mass bin $11.6 < \log_{10}(M_\star/M_\odot) < 11.7$; as expected, and shown in Figure~\ref{fig:ycyl}, the measured Compton-$y$ signal decreases toward lower stellar masses, meaning the margin between the spurious signal and the true signal narrows at lower masses. Nevertheless, in all mass bins where we have a significant detection of Compton-$y$, the spurious contribution from radio source contamination remains subdominant and is in fact comparable to the overall noise level. This empirical test strengthens our confidence that our results reflect intrinsic galaxy properties rather than residual radio source contamination. We note that comparable tests were performed on these needlet ILC maps in the context of a relativistic tSZ measurement by \citet{Coulton2026}, who found similar results.

\subsection{Previous Planck Results}
\label{sec:mmf}

\begin{figure}[ht!]
\centering         \includegraphics[width=0.70\textwidth]{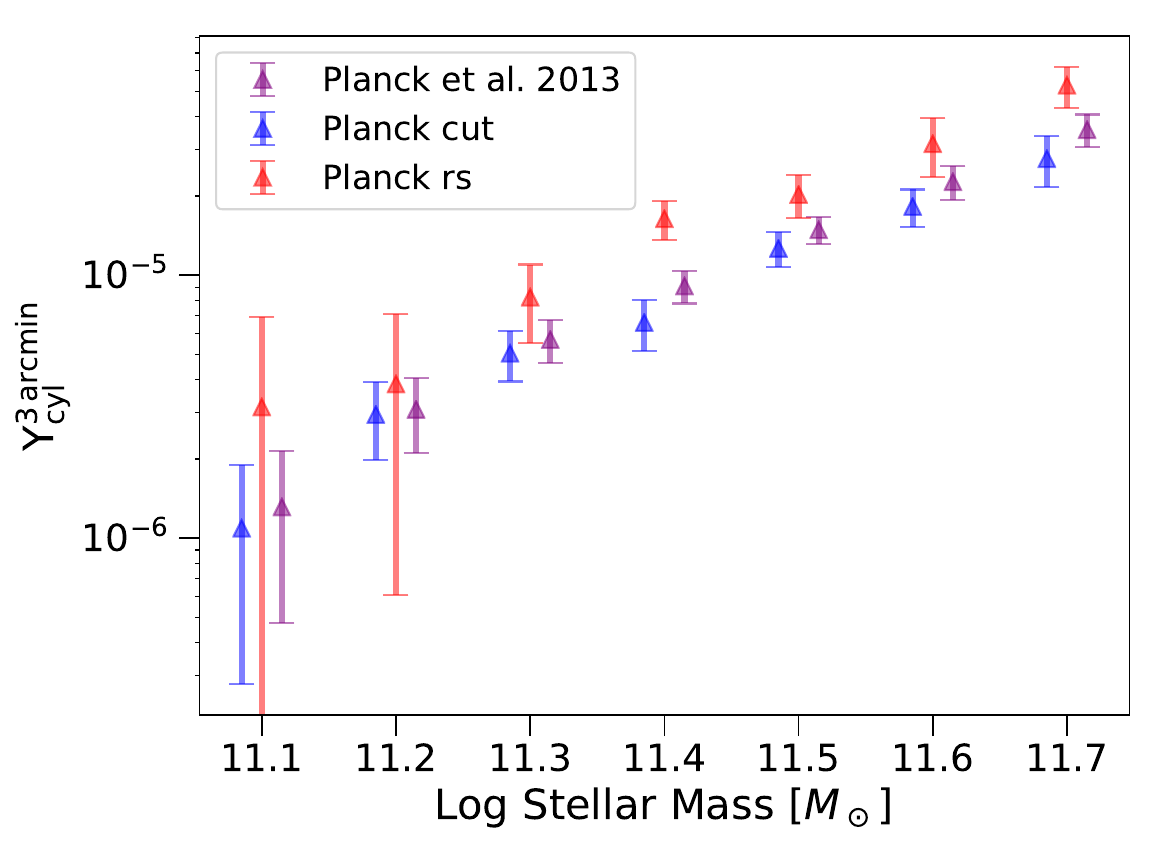}
\caption{Cylindrical Compton-$y$ signal ($Y^{3'}_{\mathrm{cyl}}$) as a 
function of stellar mass for LBGs, measured using the Planck MMF method. Results are shown for the full sample from \citet{Planck2013} (purple), and for the subsamples of galaxies without co-spatial radio sources (blue) and with co-spatial radio sources (red). Error bars denote $1\sigma$ uncertainties. For $\log_{10}(M_\star/M_\odot) > 11.4$, galaxies hosting radio sources show systematically higher Compton-$y$ signals than those without.}
\label{fig:planckrscut}
\end{figure}

In the previous LBG analysis by the Planck collaboration \citep{Planck2013}, an MMF method \citep[e.g.,][]{Melin2006} was used to extract the Compton-$y$ signal. This method is optimized to extract the Compton-$y$ signal from halos given an assumption about the pressure profile. Despite Planck's coarser angular resolution relative to ACT, the optimality of the MMF combined with a large galaxy sample size resulted in high signal-to-noise detections. These previous results also provide an independent check of our findings. The MMF method applied to Planck-only observations uses different weights than the ACT ILC maps used in our analysis, making it a complementary cross-check. We apply the same subsampling of galaxies with and without co-spatial radio sources and compute the mean MMF Compton-$y$ signal for each subsample. Figure~\ref{fig:planckrscut} shows the MMF results for each subsample alongside the results for the full sample. We find that galaxies with co-spatial radio sources have higher Compton-$y$ signals than those without for $\log_{10}(M_\star/M_\odot) > 11.4$; below this stellar mass threshold the error bars are too large to statistically differentiate between the two samples. It is likely that contamination from the central radio source contributes to these measurements, and removal of this contamination would be expected to further increase the observed differences. This result is consistent with what we find in the ACT ILC maps, but obtained with different MMF weights applied to the individual Planck frequency maps.

\subsection{Stellar Mass Distributions and CMB Halo Lensing}

\begin{figure}[ht!]
\centering \includegraphics[width=0.70\textwidth]{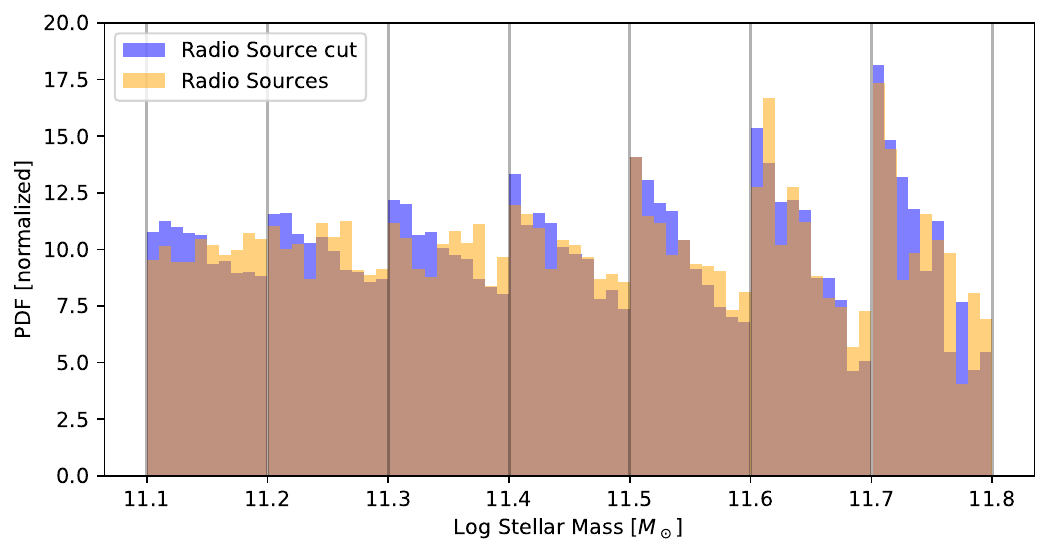}
\caption{Normalized stellar mass distributions for LBGs with co-spatial radio  sources (orange) and without (blue), spanning the range $11.1 \lesssim \log(M_*/M_\odot) \lesssim 11.8$. Each pair of histograms corresponds to a stellar mass bin of width $0.1\,\text{dex}$. The two distributions are consistent across all mass bins.}
\label{fig:smdist}
\end{figure}

The Compton-$y$ signal from a halo is a super-linear function of halo mass, hence when comparing Compton-$y$ signals between galaxy samples it is critical to account for possible mass differences. For the LBG sample, only stellar masses are publicly available; no halo masses are provided.  The stellar masses for the LBGs were provided in the NYU-VAGC \citep{Blanton2005}. We first check that the stellar mass distributions of both LBG subsamples, galaxies with and without co-spatial radio sources, are consistent with one another. Figure~\ref{fig:smdist} shows that the stellar mass distributions are practically identical. There is no evidence that the stellar mass distributions between the two samples are different. However, at fixed stellar mass it has been shown that galaxies hosting radio sources are systematically more massive in terms of halo mass than galaxies without \citep{Mandelbaum2008}. To definitively rule out that the two samples have different halo masses, and therefore different Compton-$y$ signals, we need to constrain their halo masses directly.

\begin{figure}[ht!]
\centering \includegraphics[width=0.70\textwidth]{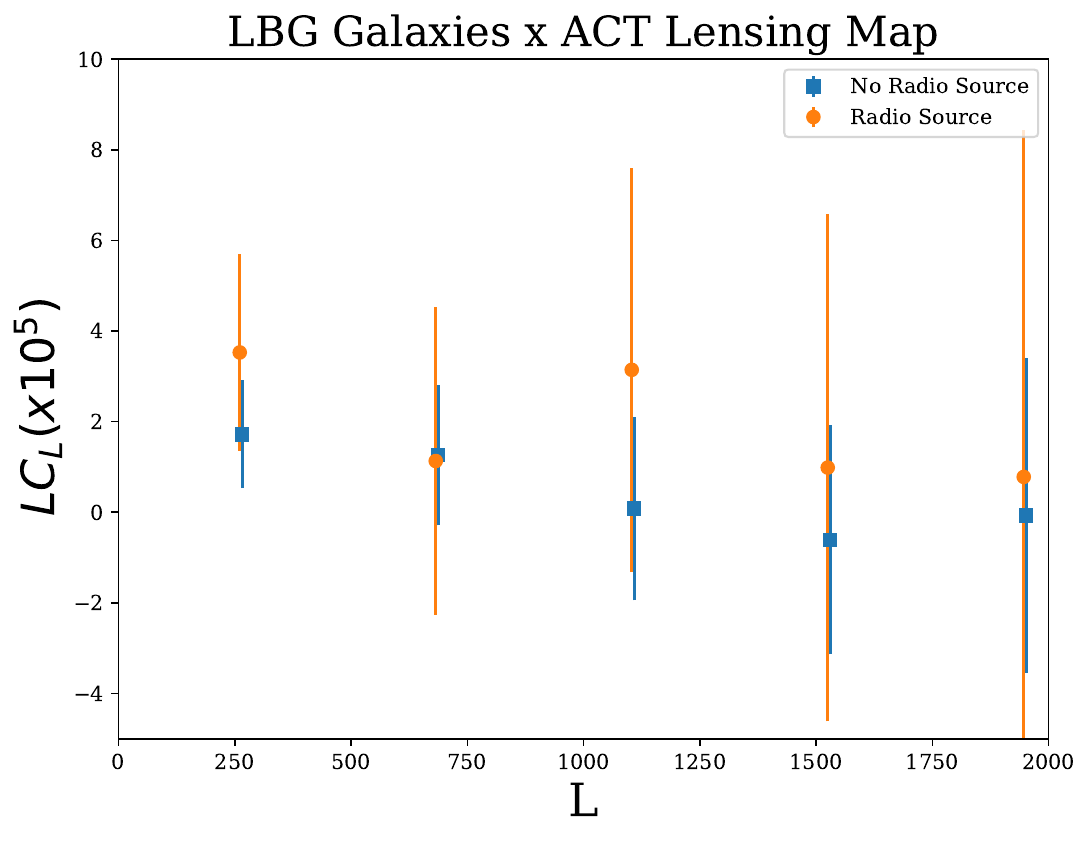}
\caption{CMB lensing convergence cross-spectrum, $LC_\ell^{\kappa g}$ ($\times 10^5$), as a function of multipole $L$ for LBGs with co-spatial radio sources (orange circles) and without (blue squares). Error bars denote the $1\sigma$ uncertainties. The two samples are consistent within the uncertainties across all multipole bins.}
\label{fig:CMBlens}
\end{figure}

We use the ACT DR6 CMB lensing maps, an additional public data product provided by ACT, to attempt to constrain the halo masses of the LBG subsamples. The ACT lensing analysis by \citet{Madhavacheril2024,Qu2024} includes a map of the CMB lensing convergence, $\kappa$, which we cross-correlate with the LBG samples. The goal is to measure the small-scale CMB lensing signal for the purpose of comparing the $\kappa$ profiles between galaxies with co-spatial radio sources and those without. This cross-correlation is not the optimal approach for small-scale CMB lensing measurements; a more optimal small-scale lensing reconstruction estimator \citep[e.g.,][]{Hu2007} is beyond the scope of this work and not guaranteed to yield more conclusive results. The cross-correlation measurement follows the methods outlined in \citet{Sherwin2012} and \citet{Allison2015}, and we compute the angular cross-spectrum ($C_\ell^{\kappa g}$) between the $\kappa$ map and galaxy overdensity maps, $g_i = \frac{n_i}{\bar{n}} - 1$, where $n_i$ is the number of sources in each pixel and $\bar{n}$ is the mean number of galaxies per pixel. The overall amplitude of $C_\ell^{\kappa g}$ is sensitive to the galaxy bias of the sample  \citep[e.g.,][]{Sherwin2012,Allison2015}, which serves as a probe of halo mass. Figure~\ref{fig:CMBlens} shows $C_\ell^{\kappa g}$ for LBGs with and without co-spatial radio sources. The measurements have a signal-to-noise ratio of around 2, and the LBGs with co-spatial radio sources show a marginally higher $C_\ell^{\kappa g}$ amplitude. However, given the large uncertainties both measurements are statistically consistent with each other, and we cannot conclusively determine whether there is a difference in halo masses between the two subsamples.

\bibliography{ref.bib}

@ARTICLE{Coulton24,
       author = {{Coulton}, William and {Madhavacheril}, Mathew S. and {Duivenvoorden}, Adriaan J. and {Hill}, J. Colin and {Abril-Cabezas}, Irene and {Ade}, Peter A.~R. and {Aiola}, Simone and {Alford}, Tommy and {Amiri}, Mandana and {Amodeo}, Stefania and {An}, Rui and {Atkins}, Zachary and {Austermann}, Jason E. and {Battaglia}, Nicholas and {Battistelli}, Elia Stefano and {Beall}, James A. and {Bean}, Rachel and {Beringue}, Benjamin and {Bhandarkar}, Tanay and {Biermann}, Emily and {Bolliet}, Boris and {Bond}, J. Richard and {Cai}, Hongbo and {Calabrese}, Erminia and {Calafut}, Victoria and {Capalbo}, Valentina and {Carrero}, Felipe and {Chesmore}, Grace E. and {Cho}, Hsiao-mei and {Choi}, Steve K. and {Clark}, Susan E. and {Rosado}, Rodrigo C{\'o}rdova and {Cothard}, Nicholas F. and {Coughlin}, Kevin and {Crowley}, Kevin T. and {Devlin}, Mark J. and {Dicker}, Simon and {Doze}, Peter and {Duell}, Cody J. and {Duff}, Shannon M. and {Dunkley}, Jo and {D{\"u}nner}, Rolando and {Fanfani}, Valentina and {Fankhanel}, Max and {Farren}, Gerrit and {Ferraro}, Simone and {Freundt}, Rodrigo and {Fuzia}, Brittany and {Gallardo}, Patricio A. and {Garrido}, Xavier and {Givans}, Jahmour and {Gluscevic}, Vera and {Golec}, Joseph E. and {Guan}, Yilun and {Halpern}, Mark and {Han}, Dongwon and {Hasselfield}, Matthew and {Healy}, Erin and {Henderson}, Shawn and {Hensley}, Brandon and {Herv{\'\i}as-Caimapo}, Carlos and {Hilton}, Gene C. and {Hilton}, Matt and {Hincks}, Adam D. and {Hlo{\v{z}}ek}, Ren{\'e}e and {Ho}, Shuay-Pwu Patty and {Huber}, Zachary B. and {Hubmayr}, Johannes and {Huffenberger}, Kevin M. and {Hughes}, John P. and {Irwin}, Kent and {Isopi}, Giovanni and {Jense}, Hidde T. and {Keller}, Ben and {Kim}, Joshua and {Knowles}, Kenda and {Koopman}, Brian J. and {Kosowsky}, Arthur and {Kramer}, Darby and {Kusiak}, Aleksandra and {La Posta}, Adrien and {Lakey}, Victoria and {Lee}, Eunseong and {Li}, Zack and {Li}, Yaqiong and {Limon}, Michele and {Lokken}, Martine and {Louis}, Thibaut and {Lungu}, Marius and {MacCrann}, Niall and {MacInnis}, Amanda and {Maldonado}, Diego and {Maldonado}, Felipe and {Mallaby-Kay}, Maya and {Marques}, Gabriela A. and {van Marrewijk}, Joshiwa and {McCarthy}, Fiona and {McMahon}, Jeff and {Mehta}, Yogesh and {Menanteau}, Felipe and {Moodley}, Kavilan and {Morris}, Thomas W. and {Mroczkowski}, Tony and {Naess}, Sigurd and {Namikawa}, Toshiya and {Nati}, Federico and {Newburgh}, Laura and {Nicola}, Andrina and {Niemack}, Michael D. and {Nolta}, Michael R. and {Orlowski-Scherer}, John and {Page}, Lyman A. and {Pandey}, Shivam and {Partridge}, Bruce and {Prince}, Heather and {Puddu}, Roberto and {Qu}, Frank J. and {Radiconi}, Federico and {Robertson}, Naomi and {Rojas}, Felipe and {Sakuma}, Tai and {Salatino}, Maria and {Schaan}, Emmanuel and {Schmitt}, Benjamin L. and {Sehgal}, Neelima and {Shaikh}, Shabbir and {Sherwin}, Blake D. and {Sierra}, Carlos and {Sievers}, Jon and {Sif{\'o}n}, Crist{\'o}bal and {Simon}, Sara and {Sonka}, Rita and {Spergel}, David N. and {Staggs}, Suzanne T. and {Storer}, Emilie and {Switzer}, Eric R. and {Tampier}, Niklas and {Thornton}, Robert and {Trac}, Hy and {Treu}, Jesse and {Tucker}, Carole and {Ullom}, Joel and {Vale}, Leila R. and {Van Engelen}, Alexander and {Van Lanen}, Jeff and {Vargas}, Cristian and {Vavagiakis}, Eve M. and {Wagoner}, Kasey and {Wang}, Yuhan and {Wenzl}, Lukas and {Wollack}, Edward J. and {Xu}, Zhilei and {Zago}, Fernando and {Zheng}, Kaiwen},
        title = "{Atacama Cosmology Telescope: High-resolution component-separated maps across one third of the sky}",
      journal = {\prd},
         year = 2024,
        month = mar,
       volume = {109},
       number = {6},
          eid = {063530},
        pages = {063530},
          doi = {10.1103/PhysRevD.109.063530},
archivePrefix = {arXiv},
       eprint = {2307.01258},
 primaryClass = {astro-ph.CO},
       adsurl = {https://ui.adsabs.harvard.edu/abs/2024PhRvD.109f3530C}
}

@ARTICLE{Naess25,
       author = {{Naess}, Sigurd and {Guan}, Yilun and {Duivenvoorden}, Adriaan J. and {Hasselfield}, Matthew and {Wang}, Yuhan and {Abril-Cabezas}, Irene and {Addison}, Graeme E. and {Ade}, Peter A.~R. and {Aiola}, Simone and {Alford}, Tommy and {Alonso}, David and {Amiri}, Mandana and {An}, Rui and {Atkins}, Zachary and {Austermann}, Jason E. and {Barbavara}, Eleonora and {Battaglia}, Nicholas and {Battistelli}, Elia Stefano and {Beall}, James A. and {Bean}, Rachel and {Beheshti}, Ali and {Beringue}, Benjamin and {Bhandarkar}, Tanay and {Biermann}, Emily and {Bolliet}, Boris and {Bond}, J Richard and {Calabrese}, Erminia and {Capalbo}, Valentina and {Carrero}, Felipe and {Chen}, Stephen and {Chesmore}, Grace and {Cho}, Hsiao-mei and {Choi}, Steve K. and {Clark}, Susan E. and {Cordova Rosado}, Rodrigo and {Cothard}, Nicholas F. and {Coughlin}, Kevin and {Coulton}, William and {Crichton}, Devin and {Crowley}, Kevin T. and {Devlin}, Mark J. and {Dicker}, Simon and {Duell}, Cody J. and {Duff}, Shannon M. and {Dunkley}, Jo and {Dunner}, Rolando and {Embil Villagra}, Carmen and {Fankhanel}, Max and {Farren}, Gerrit S. and {Ferraro}, Simone and {Foster}, Allen and {Freundt}, Rodrigo and {Fuzia}, Brittany and {Gallardo}, Patricio A. and {Garrido}, Xavier and {Giardiello}, Serena and {Gill}, Ajay and {Givans}, Jahmour and {Gluscevic}, Vera and {Golec}, Joseph E. and {Gong}, Yulin and {Halpern}, Mark and {Harrison}, Ian and {Healy}, Erin and {Henderson}, Shawn and {Hensley}, Brandon and {Herv{\'\i}as-Caimapo}, Carlos and {Hill}, J. Colin and {Hilton}, Gene C. and {Hilton}, Matt and {Hincks}, Adam D. and {Hlo{\v{z}}ek}, Ren{\'e}e and {Ho}, Shuay-Pwu Patty and {Hood}, John and {Hornecker}, Erika and {Huber}, Zachary B. and {Hubmayr}, Johannes and {Huffenberger}, Kevin M. and {Hughes}, John P. and {Ikape}, Margaret and {Irwin}, Kent and {Isopi}, Giovanni and {Jense}, Hidde T. and {Joshi}, Neha and {Keller}, Ben and {Kim}, Joshua and {Knowles}, Kenda and {Koopman}, Brian J. and {Kosowsky}, Arthur and {Kramer}, Darby and {Kusiak}, Aleksandra and {La Posta}, Adrien and {Lagu{\"e}}, Alex and {Lakey}, Victoria and {Lee}, Eunseong and {Li}, Yaqiong and {Li}, Zack and {Limon}, Michele and {Lokken}, Martine and {Louis}, Thibaut and {Lungu}, Marius and {MacCrann}, Niall and {MacInnis}, Amanda and {Madhavacheril}, Mathew S. and {Maldonado}, Diego and {Maldonado}, Felipe and {Mallaby-Kay}, Maya and {Marques}, Gabriela A. and {van Marrewijk}, Joshiwa and {McCarthy}, Fiona and {McMahon}, Jeff and {Mehta}, Yogesh and {Menanteau}, Felipe and {Moodley}, Kavilan and {Morris}, Thomas W. and {Mroczkowski}, Tony and {Namikawa}, Toshiya and {Nati}, Federico and {Nerval}, Simran K. and {Newburgh}, Laura and {Nicola}, Andrina and {Niemack}, Michael D. and {Nolta}, Michael R. and {Orlowski-Scherer}, John and {Page}, Lyman A. and {Pandey}, Shivam and {Partridge}, Bruce and {Perez Sarmiento}, Karen and {Prince}, Heather and {Puddu}, Roberto and {Qu}, Frank J. and {Quiroga}, Rodrigo and {Ragavan}, Damien C. and {Ried Guachalla}, Bernardita and {Rogers}, Keir K. and {Rojas}, Felipe and {Sakuma}, Tai and {Schaan}, Emmanuel and {Schmitt}, Benjamin L. and {Sehgal}, Neelima and {Shaikh}, Shabbir and {Sherwin}, Blake D. and {Sierra}, Carlos and {Sievers}, Jon and {Sif{\'o}n}, Crist{\'o}bal and {Simon}, Sara and {Sonka}, Rita and {Spergel}, David N. and {Staggs}, Suzanne T. and {Storer}, Emilie and {Surrao}, Kristen and {Switzer}, Eric R. and {Tampier}, Niklas and {Thornton}, Robert and {Trac}, Hy and {Tucker}, Carole and {Ullom}, Joel and {Vale}, Leila R. and {Van Engelen}, Alexander and {Van Lanen}, Jeff and {Vargas}, Cristian and {Vavagiakis}, Eve M. and {Wagoner}, Kasey and {Wenzl}, Lukas and {Wollack}, Edward J. and {Zheng}, Kaiwen},
        title = "{The Atacama Cosmology Telescope: DR6 Maps}",
      journal = {arXiv e-prints},
         year = 2025,
        month = mar,
          eid = {arXiv:2503.14451},
        pages = {arXiv:2503.14451},
          doi = {10.48550/arXiv.2503.14451},
archivePrefix = {arXiv},
       eprint = {2503.14451},
 primaryClass = {astro-ph.CO},
       adsurl = {https://ui.adsabs.harvard.edu/abs/2025arXiv250314451N}
}

@ARTICLE{Planck2020,
       author = {{Planck Collaboration} and {Akrami}, Y. and {Andersen}, K.~J. and {Ashdown}, M. and {Baccigalupi}, C. and {Ballardini}, M. and {Banday}, A.~J. and {Barreiro}, R.~B. and {Bartolo}, N. and {Basak}, S. and {Benabed}, K. and {Bernard}, J. -P. and {Bersanelli}, M. and {Bielewicz}, P. and {Bond}, J.~R. and {Borrill}, J. and {Burigana}, C. and {Butler}, R.~C. and {Calabrese}, E. and {Casaponsa}, B. and {Chiang}, H.~C. and {Colombo}, L.~P.~L. and {Combet}, C. and {Crill}, B.~P. and {Cuttaia}, F. and {de Bernardis}, P. and {de Rosa}, A. and {de Zotti}, G. and {Delabrouille}, J. and {Di Valentino}, E. and {Diego}, J.~M. and {Dor{\'e}}, O. and {Douspis}, M. and {Dupac}, X. and {Eriksen}, H.~K. and {Fernandez-Cobos}, R. and {Finelli}, F. and {Frailis}, M. and {Fraisse}, A.~A. and {Franceschi}, E. and {Frolov}, A. and {Galeotta}, S. and {Galli}, S. and {Ganga}, K. and {Gerbino}, M. and {Ghosh}, T. and {Gonz{\'a}lez-Nuevo}, J. and {G{\'o}rski}, K.~M. and {Gruppuso}, A. and {Gudmundsson}, J.~E. and {Handley}, W. and {Helou}, G. and {Herranz}, D. and {Hildebrandt}, S.~R. and {Hivon}, E. and {Huang}, Z. and {Jaffe}, A.~H. and {Jones}, W.~C. and {Keih{\"a}nen}, E. and {Keskitalo}, R. and {Kiiveri}, K. and {Kim}, J. and {Kisner}, T.~S. and {Krachmalnicoff}, N. and {Kunz}, M. and {Kurki-Suonio}, H. and {Lasenby}, A. and {Lattanzi}, M. and {Lawrence}, C.~R. and {Le Jeune}, M. and {Levrier}, F. and {Liguori}, M. and {Lilje}, P.~B. and {Lilley}, M. and {Lindholm}, V. and {L{\'o}pez-Caniego}, M. and {Lubin}, P.~M. and {Mac{\'\i}as-P{\'e}rez}, J.~F. and {Maino}, D. and {Mandolesi}, N. and {Marcos-Caballero}, A. and {Maris}, M. and {Martin}, P.~G. and {Mart{\'\i}nez-Gonz{\'a}lez}, E. and {Matarrese}, S. and {Mauri}, N. and {McEwen}, J.~D. and {Meinhold}, P.~R. and {Mennella}, A. and {Migliaccio}, M. and {Mitra}, S. and {Molinari}, D. and {Montier}, L. and {Morgante}, G. and {Moss}, A. and {Natoli}, P. and {Paoletti}, D. and {Partridge}, B. and {Patanchon}, G. and {Pearson}, D. and {Pearson}, T.~J. and {Perrotta}, F. and {Piacentini}, F. and {Polenta}, G. and {Rachen}, J.~P. and {Reinecke}, M. and {Remazeilles}, M. and {Renzi}, A. and {Rocha}, G. and {Rosset}, C. and {Roudier}, G. and {Rubi{\~n}o-Mart{\'\i}n}, J.~A. and {Ruiz-Granados}, B. and {Salvati}, L. and {Savelainen}, M. and {Scott}, D. and {Sirignano}, C. and {Sirri}, G. and {Spencer}, L.~D. and {Suur-Uski}, A. -S. and {Svalheim}, L.~T. and {Tauber}, J.~A. and {Tavagnacco}, D. and {Tenti}, M. and {Terenzi}, L. and {Thommesen}, H. and {Toffolatti}, L. and {Tomasi}, M. and {Tristram}, M. and {Trombetti}, T. and {Valiviita}, J. and {Van Tent}, B. and {Vielva}, P. and {Villa}, F. and {Vittorio}, N. and {Wandelt}, B.~D. and {Wehus}, I.~K. and {Zacchei}, A. and {Zonca}, A.},
        title = "{Planck intermediate results. LVII. Joint Planck LFI and HFI data processing}",
      journal = {\aap},
         year = 2020,
        month = nov,
       volume = {643},
          eid = {A42},
        pages = {A42},
          doi = {10.1051/0004-6361/202038073},
archivePrefix = {arXiv},
       eprint = {2007.04997},
 primaryClass = {astro-ph.CO},
       adsurl = {https://ui.adsabs.harvard.edu/abs/2020A&A...643A..42P}
}

@ARTICLE{Crichton2016,
       author = {{Crichton}, Devin and {Gralla}, Megan B. and {Hall}, Kirsten and {Marriage}, Tobias A. and {Zakamska}, Nadia L. and {Battaglia}, Nick and {Bond}, J. Richard and {Devlin}, Mark J. and {Hill}, J. Colin and {Hilton}, Matt and {Hincks}, Adam D. and {Huffenberger}, Kevin M. and {Hughes}, John P. and {Kosowsky}, Arthur and {Moodley}, Kavilan and {Niemack}, Michael D. and {Page}, Lyman A. and {Partridge}, Bruce and {Sievers}, Jonathan L. and {Sif{\'o}n}, Crist{\'o}bal and {Staggs}, Suzanne T. and {Viero}, Marco P. and {Wollack}, Edward J.},
        title = "{Evidence for the thermal Sunyaev-Zel'dovich effect associated with quasar feedback}",
      journal = {\mnras},
         year = 2016,
        month = may,
       volume = {458},
       number = {2},
        pages = {1478-1492},
          doi = {10.1093/mnras/stw344},
archivePrefix = {arXiv},
       eprint = {1510.05656},
 primaryClass = {astro-ph.GA},
       adsurl = {https://ui.adsabs.harvard.edu/abs/2016MNRAS.458.1478C}
}

@ARTICLE{Aiola2020,
       author = {{Aiola}, Simone and {Calabrese}, Erminia and {Maurin}, Lo{\"\i}c and {Naess}, Sigurd and {Schmitt}, Benjamin L. and {Abitbol}, Maximilian H. and {Addison}, Graeme E. and {Ade}, Peter A.~R. and {Alonso}, David and {Amiri}, Mandana and {Amodeo}, Stefania and {Angile}, Elio and {Austermann}, Jason E. and {Baildon}, Taylor and {Battaglia}, Nick and {Beall}, James A. and {Bean}, Rachel and {Becker}, Daniel T. and {Bond}, J. Richard and {Bruno}, Sarah Marie and {Calafut}, Victoria and {Campusano}, Luis E. and {Carrero}, Felipe and {Chesmore}, Grace E. and {Cho}, Hsiao-mei and {Choi}, Steve K. and {Clark}, Susan E. and {Cothard}, Nicholas F. and {Crichton}, Devin and {Crowley}, Kevin T. and {Darwish}, Omar and {Datta}, Rahul and {Denison}, Edward V. and {Devlin}, Mark J. and {Duell}, Cody J. and {Duff}, Shannon M. and {Duivenvoorden}, Adriaan J. and {Dunkley}, Jo and {D{\"u}nner}, Rolando and {Essinger-Hileman}, Thomas and {Fankhanel}, Max and {Ferraro}, Simone and {Fox}, Anna E. and {Fuzia}, Brittany and {Gallardo}, Patricio A. and {Gluscevic}, Vera and {Golec}, Joseph E. and {Grace}, Emily and {Gralla}, Megan and {Guan}, Yilun and {Hall}, Kirsten and {Halpern}, Mark and {Han}, Dongwon and {Hargrave}, Peter and {Hasselfield}, Matthew and {Helton}, Jakob M. and {Henderson}, Shawn and {Hensley}, Brandon and {Hill}, J. Colin and {Hilton}, Gene C. and {Hilton}, Matt and {Hincks}, Adam D. and {Hlo{\v{z}}ek}, Ren{\'e}e and {Ho}, Shuay-Pwu Patty and {Hubmayr}, Johannes and {Huffenberger}, Kevin M. and {Hughes}, John P. and {Infante}, Leopoldo and {Irwin}, Kent and {Jackson}, Rebecca and {Klein}, Jeff and {Knowles}, Kenda and {Koopman}, Brian and {Kosowsky}, Arthur and {Lakey}, Vincent and {Li}, Dale and {Li}, Yaqiong and {Li}, Zack and {Lokken}, Martine and {Louis}, Thibaut and {Lungu}, Marius and {MacInnis}, Amanda and {Madhavacheril}, Mathew and {Maldonado}, Felipe and {Mallaby-Kay}, Maya and {Marsden}, Danica and {McMahon}, Jeff and {Menanteau}, Felipe and {Moodley}, Kavilan and {Morton}, Tim and {Namikawa}, Toshiya and {Nati}, Federico and {Newburgh}, Laura and {Nibarger}, John P. and {Nicola}, Andrina and {Niemack}, Michael D. and {Nolta}, Michael R. and {Orlowski-Sherer}, John and {Page}, Lyman A. and {Pappas}, Christine G. and {Partridge}, Bruce and {Phakathi}, Phumlani and {Pisano}, Giampaolo and {Prince}, Heather and {Puddu}, Roberto and {Qu}, Frank J. and {Rivera}, Jesus and {Robertson}, Naomi and {Rojas}, Felipe and {Salatino}, Maria and {Schaan}, Emmanuel and {Schillaci}, Alessandro and {Sehgal}, Neelima and {Sherwin}, Blake D. and {Sierra}, Carlos and {Sievers}, Jon and {Sifon}, Cristobal and {Sikhosana}, Precious and {Simon}, Sara and {Spergel}, David N. and {Staggs}, Suzanne T. and {Stevens}, Jason and {Storer}, Emilie and {Sunder}, Dhaneshwar D. and {Switzer}, Eric R. and {Thorne}, Ben and {Thornton}, Robert and {Trac}, Hy and {Treu}, Jesse and {Tucker}, Carole and {Vale}, Leila R. and {Van Engelen}, Alexander and {Van Lanen}, Jeff and {Vavagiakis}, Eve M. and {Wagoner}, Kasey and {Wang}, Yuhan and {Ward}, Jonathan T. and {Wollack}, Edward J. and {Xu}, Zhilei and {Zago}, Fernando and {Zhu}, Ningfeng},
        title = "{The Atacama Cosmology Telescope: DR4 maps and cosmological parameters}",
      journal = {\jcap},
         year = 2020,
        month = dec,
       volume = {2020},
       number = {12},
          eid = {047},
        pages = {047},
          doi = {10.1088/1475-7516/2020/12/047},
archivePrefix = {arXiv},
       eprint = {2007.07288},
 primaryClass = {astro-ph.CO},
       adsurl = {https://ui.adsabs.harvard.edu/abs/2020JCAP...12..047A}
}

@ARTICLE{Greco2015,
       author = {{Greco}, Johnny P. and {Hill}, J. Colin and {Spergel}, David N. and {Battaglia}, Nicholas},
        title = "{The Stacked Thermal Sunyaev-Zel'dovich Signal of Locally Brightest Galaxies in Planck Full Mission Data: Evidence for Galaxy Feedback?}",
      journal = {\apj},
         year = 2015,
        month = aug,
       volume = {808},
       number = {2},
          eid = {151},
        pages = {151},
          doi = {10.1088/0004-637X/808/2/151},
archivePrefix = {arXiv},
       eprint = {1409.6747},
 primaryClass = {astro-ph.CO},
       adsurl = {https://ui.adsabs.harvard.edu/abs/2015ApJ...808..151G}
}

@ARTICLE{Planck2013,
       author = {{Planck Collaboration} and {Ade}, P.~A.~R. and {Aghanim}, N. and {Arnaud}, M. and {Ashdown}, M. and {Atrio-Barandela}, F. and {Aumont}, J. and {Baccigalupi}, C. and {Balbi}, A. and {Banday}, A.~J. and {Barreiro}, R.~B. and {Barrena}, R. and {Bartlett}, J.~G. and {Battaner}, E. and {Benabed}, K. and {Bernard}, J. -P. and {Bersanelli}, M. and {Bikmaev}, I. and {Bock}, J.~J. and {B{\"o}hringer}, H. and {Bonaldi}, A. and {Bond}, J.~R. and {Borrill}, J. and {Bouchet}, F.~R. and {Bourdin}, H. and {Burenin}, R. and {Burigana}, C. and {Butler}, R.~C. and {Cabella}, P. and {Chamballu}, A. and {Chary}, R. -R. and {Chiang}, L. -Y. and {Chon}, G. and {Christensen}, P.~R. and {Clements}, D.~L. and {Colafrancesco}, S. and {Colombi}, S. and {Colombo}, L.~P.~L. and {Comis}, B. and {Coulais}, A. and {Crill}, B.~P. and {Cuttaia}, F. and {Da Silva}, A. and {Dahle}, H. and {Davis}, R.~J. and {de Bernardis}, P. and {de Gasperis}, G. and {de Rosa}, A. and {de Zotti}, G. and {Delabrouille}, J. and {D{\'e}mocl{\`e}s}, J. and {Diego}, J.~M. and {Dole}, H. and {Donzelli}, S. and {Dor{\'e}}, O. and {Douspis}, M. and {Dupac}, X. and {Efstathiou}, G. and {En{\ss}lin}, T.~A. and {Finelli}, F. and {Flores-Cacho}, I. and {Forni}, O. and {Frailis}, M. and {Franceschi}, E. and {Frommert}, M. and {Galeotta}, S. and {Ganga}, K. and {G{\'e}nova-Santos}, R.~T. and {Giard}, M. and {Giraud-H{\'e}raud}, Y. and {Gonz{\'a}lez-Nuevo}, J. and {G{\'o}rski}, K.~M. and {Gregorio}, A. and {Gruppuso}, A. and {Hansen}, F.~K. and {Harrison}, D. and {Hern{\'a}ndez-Monteagudo}, C. and {Herranz}, D. and {Hildebrandt}, S.~R. and {Hivon}, E. and {Hobson}, M. and {Holmes}, W.~A. and {Hornstrup}, A. and {Hovest}, W. and {Huffenberger}, K.~M. and {Hurier}, G. and {Jaffe}, T.~R. and {Jaffe}, A.~H. and {Jones}, W.~C. and {Juvela}, M. and {Keih{\"a}nen}, E. and {Keskitalo}, R. and {Khamitov}, I. and {Kisner}, T.~S. and {Kneissl}, R. and {Knoche}, J. and {Kunz}, M. and {Kurki-Suonio}, H. and {L{\"a}hteenm{\"a}ki}, A. and {Lamarre}, J. -M. and {Lasenby}, A. and {Lawrence}, C.~R. and {Le Jeune}, M. and {Leonardi}, R. and {Lilje}, P.~B. and {Linden-V{\o}rnle}, M. and {L{\'o}pez-Caniego}, M. and {Lubin}, P.~M. and {Luzzi}, G. and {Mac{\'\i}as-P{\'e}rez}, J.~F. and {MacTavish}, C.~J. and {Maffei}, B. and {Maino}, D. and {Mandolesi}, N. and {Maris}, M. and {Marleau}, F. and {Marshall}, D.~J. and {Mart{\'\i}nez-Gonz{\'a}lez}, E. and {Masi}, S. and {Massardi}, M. and {Matarrese}, S. and {Mazzotta}, P. and {Mei}, S. and {Melchiorri}, A. and {Melin}, J. -B. and {Mendes}, L. and {Mennella}, A. and {Mitra}, S. and {Miville-Desch{\^e}nes}, M. -A. and {Moneti}, A. and {Montier}, L. and {Morgante}, G. and {Mortlock}, D. and {Munshi}, D. and {Murphy}, J.~A. and {Naselsky}, P. and {Nati}, F. and {Natoli}, P. and {N{\o}rgaard-Nielsen}, H.~U. and {Noviello}, F. and {Novikov}, D. and {Novikov}, I. and {Osborne}, S. and {Oxborrow}, C.~A. and {Pajot}, F. and {Paoletti}, D. and {Perotto}, L. and {Perrotta}, F. and {Piacentini}, F. and {Piat}, M. and {Pierpaoli}, E. and {Piffaretti}, R. and {Plaszczynski}, S. and {Pointecouteau}, E. and {Polenta}, G. and {Popa}, L. and {Poutanen}, T. and {Pratt}, G.~W. and {Prunet}, S. and {Puget}, J. -L. and {Rachen}, J.~P. and {Rebolo}, R. and {Reinecke}, M. and {Remazeilles}, M. and {Renault}, C. and {Ricciardi}, S. and {Ristorcelli}, I. and {Rocha}, G. and {Roman}, M. and {Rosset}, C. and {Rossetti}, M. and {Rubi{\~n}o-Mart{\'\i}n}, J.~A. and {Rusholme}, B. and {Sandri}, M. and {Savini}, G. and {Scott}, D. and {Spencer}, L. and {Starck}, J. -L. and {Stolyarov}, V. and {Sudiwala}, R. and {Sunyaev}, R. and {Sutton}, D. and {Suur-Uski}, A. -S. and {Sygnet}, J. -F. and {Tauber}, J.~A. and {Terenzi}, L. and {Toffolatti}, L. and {Tomasi}, M. and {Tristram}, M. and {Valenziano}, L. and {Van Tent}, B. and {Vielva}, P. and {Villa}, F. and {Vittorio}, N. and {Wade}, L.~A. and {Wandelt}, B.~D. and {Wang}, W. and {Welikala}, N. and {Weller}, J. and {White}, S.~D.~M.},
        title = "{Planck intermediate results. XI. The gas content of dark matter halos: the Sunyaev-Zeldovich-stellar mass relation for locally brightest galaxies}",
      journal = {\aap},
         year = 2013,
        month = sep,
       volume = {557},
          eid = {A52},
        pages = {A52},
          doi = {10.1051/0004-6361/201220941},
archivePrefix = {arXiv},
       eprint = {1212.4131},
 primaryClass = {astro-ph.CO},
       adsurl = {https://ui.adsabs.harvard.edu/abs/2013A&A...557A..52P}
}

@ARTICLE{SDSSDR7,
       author = {{Abazajian}, Kevork N. and {Adelman-McCarthy}, Jennifer K. and {Ag{\"u}eros}, Marcel A. and {Allam}, Sahar S. and {Allende Prieto}, Carlos and {An}, Deokkeun and {Anderson}, Kurt S.~J. and {Anderson}, Scott F. and {Annis}, James and {Bahcall}, Neta A. and {Bailer-Jones}, C.~A.~L. and {Barentine}, J.~C. and {Bassett}, Bruce A. and {Becker}, Andrew C. and {Beers}, Timothy C. and {Bell}, Eric F. and {Belokurov}, Vasily and {Berlind}, Andreas A. and {Berman}, Eileen F. and {Bernardi}, Mariangela and {Bickerton}, Steven J. and {Bizyaev}, Dmitry and {Blakeslee}, John P. and {Blanton}, Michael R. and {Bochanski}, John J. and {Boroski}, William N. and {Brewington}, Howard J. and {Brinchmann}, Jarle and {Brinkmann}, J. and {Brunner}, Robert J. and {Budav{\'a}ri}, Tam{\'a}s and {Carey}, Larry N. and {Carliles}, Samuel and {Carr}, Michael A. and {Castander}, Francisco J. and {Cinabro}, David and {Connolly}, A.~J. and {Csabai}, Istv{\'a}n and {Cunha}, Carlos E. and {Czarapata}, Paul C. and {Davenport}, James R.~A. and {de Haas}, Ernst and {Dilday}, Ben and {Doi}, Mamoru and {Eisenstein}, Daniel J. and {Evans}, Michael L. and {Evans}, N.~W. and {Fan}, Xiaohui and {Friedman}, Scott D. and {Frieman}, Joshua A. and {Fukugita}, Masataka and {G{\"a}nsicke}, Boris T. and {Gates}, Evalyn and {Gillespie}, Bruce and {Gilmore}, G. and {Gonzalez}, Belinda and {Gonzalez}, Carlos F. and {Grebel}, Eva K. and {Gunn}, James E. and {Gy{\"o}ry}, Zsuzsanna and {Hall}, Patrick B. and {Harding}, Paul and {Harris}, Frederick H. and {Harvanek}, Michael and {Hawley}, Suzanne L. and {Hayes}, Jeffrey J.~E. and {Heckman}, Timothy M. and {Hendry}, John S. and {Hennessy}, Gregory S. and {Hindsley}, Robert B. and {Hoblitt}, J. and {Hogan}, Craig J. and {Hogg}, David W. and {Holtzman}, Jon A. and {Hyde}, Joseph B. and {Ichikawa}, Shin-ichi and {Ichikawa}, Takashi and {Im}, Myungshin and {Ivezi{\'c}}, {\v{Z}}eljko and {Jester}, Sebastian and {Jiang}, Linhua and {Johnson}, Jennifer A. and {Jorgensen}, Anders M. and {Juri{\'c}}, Mario and {Kent}, Stephen M. and {Kessler}, R. and {Kleinman}, S.~J. and {Knapp}, G.~R. and {Konishi}, Kohki and {Kron}, Richard G. and {Krzesinski}, Jurek and {Kuropatkin}, Nikolay and {Lampeitl}, Hubert and {Lebedeva}, Svetlana and {Lee}, Myung Gyoon and {Lee}, Young Sun and {French Leger}, R. and {L{\'e}pine}, S{\'e}bastien and {Li}, Nolan and {Lima}, Marcos and {Lin}, Huan and {Long}, Daniel C. and {Loomis}, Craig P. and {Loveday}, Jon and {Lupton}, Robert H. and {Magnier}, Eugene and {Malanushenko}, Olena and {Malanushenko}, Viktor and {Mandelbaum}, Rachel and {Margon}, Bruce and {Marriner}, John P. and {Mart{\'\i}nez-Delgado}, David and {Matsubara}, Takahiko and {McGehee}, Peregrine M. and {McKay}, Timothy A. and {Meiksin}, Avery and {Morrison}, Heather L. and {Mullally}, Fergal and {Munn}, Jeffrey A. and {Murphy}, Tara and {Nash}, Thomas and {Nebot}, Ada and {Neilsen}, Jr., Eric H. and {Newberg}, Heidi Jo and {Newman}, Peter R. and {Nichol}, Robert C. and {Nicinski}, Tom and {Nieto-Santisteban}, Maria and {Nitta}, Atsuko and {Okamura}, Sadanori and {Oravetz}, Daniel J. and {Ostriker}, Jeremiah P. and {Owen}, Russell and {Padmanabhan}, Nikhil and {Pan}, Kaike and {Park}, Changbom and {Pauls}, George and {Peoples}, Jr., John and {Percival}, Will J. and {Pier}, Jeffrey R. and {Pope}, Adrian C. and {Pourbaix}, Dimitri and {Price}, Paul A. and {Purger}, Norbert and {Quinn}, Thomas and {Raddick}, M. Jordan and {Re Fiorentin}, Paola and {Richards}, Gordon T. and {Richmond}, Michael W. and {Riess}, Adam G. and {Rix}, Hans-Walter and {Rockosi}, Constance M. and {Sako}, Masao and {Schlegel}, David J. and {Schneider}, Donald P. and {Scholz}, Ralf-Dieter and {Schreiber}, Matthias R. and {Schwope}, Axel D. and {Seljak}, Uro{\v{s}} and {Sesar}, Branimir and {Sheldon}, Erin and {Shimasaku}, Kazu and {Sibley}, Valena C. and {Simmons}, A.~E. and {Sivarani}, Thirupathi and {Allyn Smith}, J. and {Smith}, Martin C. and {Smol{\v{c}}i{\'c}}, Vernesa and {Snedden}, Stephanie A. and {Stebbins}, Albert and {Steinmetz}, Matthias and {Stoughton}, Chris and {Strauss}, Michael A. and {SubbaRao}, Mark and {Suto}, Yasushi and {Szalay}, Alexander S. and {Szapudi}, Istv{\'a}n and {Szkody}, Paula and {Tanaka}, Masayuki and {Tegmark}, Max and {Teodoro}, Luis F.~A. and {Thakar}, Aniruddha R. and {Tremonti}, Christy A. and {Tucker}, Douglas L. and {Uomoto}, Alan and {Vanden Berk}, Daniel E. and {Vandenberg}, Jan and {Vidrih}, S. and {Vogeley}, Michael S. and {Voges}, Wolfgang and {Vogt}, Nicole P. and {Wadadekar}, Yogesh and {Watters}, Shannon and {Weinberg}, David H. and {West}, Andrew A. and {White}, Simon D.~M. and {Wilhite}, Brian C. and {Wonders}, Alainna C. and {Yanny}, Brian and {Yocum}, D.~R.},
        title = "{The Seventh Data Release of the Sloan Digital Sky Survey}",
      journal = {\apjs},
         year = 2009,
        month = jun,
       volume = {182},
       number = {2},
        pages = {543-558},
          doi = {10.1088/0067-0049/182/2/543},
archivePrefix = {arXiv},
       eprint = {0812.0649},
 primaryClass = {astro-ph},
       adsurl = {https://ui.adsabs.harvard.edu/abs/2009ApJS..182..543A}
}

@ARTICLE{Blanton2005,
       author = {{Blanton}, Michael R. and {Schlegel}, David J. and {Strauss}, Michael A. and {Brinkmann}, J. and {Finkbeiner}, Douglas and {Fukugita}, Masataka and {Gunn}, James E. and {Hogg}, David W. and {Ivezi{\'c}}, {\v{Z}}eljko and {Knapp}, G.~R. and {Lupton}, Robert H. and {Munn}, Jeffrey A. and {Schneider}, Donald P. and {Tegmark}, Max and {Zehavi}, Idit},
        title = "{New York University Value-Added Galaxy Catalog: A Galaxy Catalog Based on New Public Surveys}",
      journal = {\aj},
         year = 2005,
        month = jun,
       volume = {129},
       number = {6},
        pages = {2562-2578},
          doi = {10.1086/429803},
archivePrefix = {arXiv},
       eprint = {astro-ph/0410166},
 primaryClass = {astro-ph},
       adsurl = {https://ui.adsabs.harvard.edu/abs/2005AJ....129.2562B}
}

@ARTICLE{Cunha2009,
       author = {{Cunha}, Carlos E. and {Lima}, Marcos and {Oyaizu}, Hiroaki and {Frieman}, Joshua and {Lin}, Huan},
        title = "{Estimating the redshift distribution of photometric galaxy samples - II. Applications and tests of a new method}",
      journal = {\mnras},
         year = 2009,
        month = jul,
       volume = {396},
       number = {4},
        pages = {2379-2398},
          doi = {10.1111/j.1365-2966.2009.14908.x},
archivePrefix = {arXiv},
       eprint = {0810.2991},
 primaryClass = {astro-ph},
       adsurl = {https://ui.adsabs.harvard.edu/abs/2009MNRAS.396.2379C}
}

@ARTICLE{Schaan2021,
       author = {{Schaan}, Emmanuel and {Ferraro}, Simone and {Amodeo}, Stefania and {Battaglia}, Nicholas and {Aiola}, Simone and {Austermann}, Jason E. and {Beall}, James A. and {Bean}, Rachel and {Becker}, Daniel T. and {Bond}, Richard J. and {Calabrese}, Erminia and {Calafut}, Victoria and {Choi}, Steve K. and {Denison}, Edward V. and {Devlin}, Mark J. and {Duff}, Shannon M. and {Duivenvoorden}, Adriaan J. and {Dunkley}, Jo and {D{\"u}nner}, Rolando and {Gallardo}, Patricio A. and {Guan}, Yilun and {Han}, Dongwon and {Hill}, J. Colin and {Hilton}, Gene C. and {Hilton}, Matt and {Hlo{\v{z}}ek}, Ren{\'e}e and {Hubmayr}, Johannes and {Huffenberger}, Kevin M. and {Hughes}, John P. and {Koopman}, Brian J. and {MacInnis}, Amanda and {McMahon}, Jeff and {Madhavacheril}, Mathew S. and {Moodley}, Kavilan and {Mroczkowski}, Tony and {Naess}, Sigurd and {Nati}, Federico and {Newburgh}, Laura B. and {Niemack}, Michael D. and {Page}, Lyman A. and {Partridge}, Bruce and {Salatino}, Maria and {Sehgal}, Neelima and {Schillaci}, Alessandro and {Sif{\'o}n}, Crist{\'o}bal and {Smith}, Kendrick M. and {Spergel}, David N. and {Staggs}, Suzanne and {Storer}, Emilie R. and {Trac}, Hy and {Ullom}, Joel N. and {Van Lanen}, Jeff and {Vale}, Leila R. and {van Engelen}, Alexander and {Maga{\~n}a}, Mariana Vargas and {Vavagiakis}, Eve M. and {Wollack}, Edward J. and {Xu}, Zhilei and {Atacama Cosmology Telescope Collaboration}},
        title = "{Atacama Cosmology Telescope: Combined kinematic and thermal Sunyaev-Zel'dovich measurements from BOSS CMASS and LOWZ halos}",
      journal = {\prd},
         year = 2021,
        month = mar,
       volume = {103},
       number = {6},
          eid = {063513},
        pages = {063513},
          doi = {10.1103/PhysRevD.103.063513},
archivePrefix = {arXiv},
       eprint = {2009.05557},
 primaryClass = {astro-ph.CO},
       adsurl = {https://ui.adsabs.harvard.edu/abs/2021PhRvD.103f3513S}
}

@ARTICLE{McCarthyHill2024,
       author = {{McCarthy}, Fiona and {Hill}, J. Colin},
        title = "{Component-separated, CIB-cleaned thermal Sunyaev-Zel'dovich maps from Planck PR4 data with a flexible public needlet ILC pipeline}",
      journal = {\prd},
         year = 2024,
        month = jan,
       volume = {109},
       number = {2},
          eid = {023528},
        pages = {023528},
          doi = {10.1103/PhysRevD.109.023528},
archivePrefix = {arXiv},
       eprint = {2307.01043},
 primaryClass = {astro-ph.CO},
       adsurl = {https://ui.adsabs.harvard.edu/abs/2024PhRvD.109b3528M}
}

@ARTICLE{Ruan2015,
       author = {{Ruan}, John J. and {McQuinn}, Matthew and {Anderson}, Scott F.},
        title = "{Detection of Quasar Feedback from the Thermal Sunyaev-Zel{\textquoteright}dovich Effect in Planck}",
      journal = {\apj},
         year = 2015,
        month = apr,
       volume = {802},
       number = {2},
          eid = {135},
        pages = {135},
          doi = {10.1088/0004-637X/802/2/135},
archivePrefix = {arXiv},
       eprint = {1502.01723},
 primaryClass = {astro-ph.CO},
       adsurl = {https://ui.adsabs.harvard.edu/abs/2015ApJ...802..135R}
}

@ARTICLE{Liu2025,
       author = {{Liu}, R. Henry and {Ferraro}, Simone and {Schaan}, Emmanuel and {Zhou}, Rongpu and {Aguilar}, Jessica Nicole and {Ahlen}, Steven and {Battaglia}, Nicholas and {Bianchi}, Davide and {Brooks}, David and {Claybaugh}, Todd and {Cole}, Shaun and {Coulton}, William R. and {de la Macorra}, Axel and {Dey}, Arjun and {Fanning}, Kevin and {Forero-Romero}, Jaime E. and {Gazta{\~n}aga}, Enrique and {Gong}, Yulin and {Gontcho}, Satya Gontcho A. and {Gruen}, Daniel and {Gutierrez}, Gaston and {Hadzhiyska}, Boryana and {Honscheid}, Klaus and {Howlett}, Cullan and {Kehoe}, Robert and {Kisner}, Theodore and {Kremin}, Anthony and {Kusiak}, Aleksandra and {Lambert}, Andrew and {Landriau}, Martin and {Le Guillou}, Laurent and {Levi}, Michael and {Lokken}, Martine and {Manera}, Marc and {Martini}, Paul and {Meisner}, Aaron and {Miquel}, Ramon and {Moodley}, Kavilan and {Newman}, Jeffrey A. and {Niz}, Gustavo and {Palanque-Delabrouille}, Nathalie and {Percival}, Will and {Prada}, Francisco and {P{\'e}rez-R{\`a}fols}, Ignasi and {Ried Guachalla}, Bernardita and {Rossi}, Graziano and {Sanchez}, Eusebio and {Schlegel}, David and {Schubnell}, Michael and {Seo}, Hee-Jong and {Sif{\'o}n}, Crist{\'o}bal and {Sprayberry}, David and {Tarl{\'e}}, Gregory and {Vavagiakis}, Eve M. and {Weaver}, Benjamin Alan and {Wollack}, Edward J. and {Zou}, Hu},
        title = "{Measurements of the thermal Sunyaev-Zel'dovich effect with ACT and DESI luminous red galaxies}",
      journal = {\prd},
         year = 2025,
        month = oct,
       volume = {112},
       number = {8},
          eid = {083561},
        pages = {083561},
          doi = {10.1103/jqn8-19gx},
archivePrefix = {arXiv},
       eprint = {2502.08850},
 primaryClass = {astro-ph.CO},
       adsurl = {https://ui.adsabs.harvard.edu/abs/2025PhRvD.112h3561L}
}

@ARTICLE{Hadzhiyska2025,
       author = {{Hadzhiyska}, Boryana and {Ferraro}, Simone and {Farren}, Gerrit S. and {Sailer}, Noah and {Zhou}, Rongpu},
        title = "{Missing baryons recovered: A measurement of the gas fraction in galaxies and groups with the kinematic Sunyaev-Zel'dovich effect and CMB lensing}",
      journal = {\prd},
         year = 2025,
        month = dec,
       volume = {112},
       number = {12},
          eid = {123507},
        pages = {123507},
          doi = {10.1103/mdhz-fgj8},
archivePrefix = {arXiv},
       eprint = {2507.14136},
 primaryClass = {astro-ph.CO},
       adsurl = {https://ui.adsabs.harvard.edu/abs/2025PhRvD.112l3507H}
}

@ARTICLE{Worpel2013,
       author = {{Worpel}, Hauke and {Brown}, Michael J.~I. and {Jones}, D. Heath and {Floyd}, David J.~E. and {Beutler}, Florian},
        title = "{The Clustering of Galaxies around Radio-loud Active Galactic Nuclei}",
      journal = {\apj},
         year = 2013,
        month = jul,
       volume = {772},
       number = {1},
          eid = {64},
        pages = {64},
          doi = {10.1088/0004-637X/772/1/64},
archivePrefix = {arXiv},
       eprint = {1305.2673},
 primaryClass = {astro-ph.CO},
       adsurl = {https://ui.adsabs.harvard.edu/abs/2013ApJ...772...64W}
}

@ARTICLE{BBPS1,
       author = {{Battaglia}, N. and {Bond}, J.~R. and {Pfrommer}, C. and {Sievers}, J.~L.},
        title = "{On the Cluster Physics of Sunyaev-Zel'dovich and X-Ray Surveys. I. The Influence of Feedback, Non-thermal Pressure, and Cluster Shapes on Y-M Scaling Relations}",
      journal = {\apj},
         year = 2012,
        month = oct,
       volume = {758},
       number = {2},
          eid = {74},
        pages = {74},
          doi = {10.1088/0004-637X/758/2/74},
archivePrefix = {arXiv},
       eprint = {1109.3709},
 primaryClass = {astro-ph.CO},
       adsurl = {https://ui.adsabs.harvard.edu/abs/2012ApJ...758...74B}
}

@ARTICLE{Guo2011,
       author = {{Guo}, Qi and {White}, Simon and {Boylan-Kolchin}, Michael and {De Lucia}, Gabriella and {Kauffmann}, Guinevere and {Lemson}, Gerard and {Li}, Cheng and {Springel}, Volker and {Weinmann}, Simone},
        title = "{From dwarf spheroidals to cD galaxies: simulating the galaxy population in a {\ensuremath{\Lambda}}CDM cosmology}",
      journal = {\mnras},
         year = 2011,
        month = may,
       volume = {413},
       number = {1},
        pages = {101-131},
          doi = {10.1111/j.1365-2966.2010.18114.x},
archivePrefix = {arXiv},
       eprint = {1006.0106},
 primaryClass = {astro-ph.CO},
       adsurl = {https://ui.adsabs.harvard.edu/abs/2011MNRAS.413..101G}
}

@ARTICLE{Angulo2010,
       author = {{Angulo}, R.~E. and {White}, S.~D.~M.},
        title = "{One simulation to fit them all - changing the background parameters of a cosmological N-body simulation}",
      journal = {\mnras},
         year = 2010,
        month = jun,
       volume = {405},
       number = {1},
        pages = {143-154},
          doi = {10.1111/j.1365-2966.2010.16459.x},
archivePrefix = {arXiv},
       eprint = {0912.4277},
 primaryClass = {astro-ph.CO},
       adsurl = {https://ui.adsabs.harvard.edu/abs/2010MNRAS.405..143A}
}

@ARTICLE{Springel2005,
       author = {{Springel}, Volker and {White}, Simon D.~M. and {Jenkins}, Adrian and {Frenk}, Carlos S. and {Yoshida}, Naoki and {Gao}, Liang and {Navarro}, Julio and {Thacker}, Robert and {Croton}, Darren and {Helly}, John and {Peacock}, John A. and {Cole}, Shaun and {Thomas}, Peter and {Couchman}, Hugh and {Evrard}, August and {Colberg}, J{\"o}rg and {Pearce}, Frazer},
        title = "{Simulations of the formation, evolution and clustering of galaxies and quasars}",
      journal = {\nat},
         year = 2005,
        month = jun,
       volume = {435},
       number = {7042},
        pages = {629-636},
          doi = {10.1038/nature03597},
archivePrefix = {arXiv},
       eprint = {astro-ph/0504097},
 primaryClass = {astro-ph},
       adsurl = {https://ui.adsabs.harvard.edu/abs/2005Natur.435..629S}
}

@ARTICLE{Narayan1994,
       author = {{Narayan}, Ramesh and {Yi}, Insu},
        title = "{Advection-dominated Accretion: A Self-similar Solution}",
      journal = {\apjl},
         year = 1994,
        month = jun,
       volume = {428},
        pages = {L13},
          doi = {10.1086/187381},
archivePrefix = {arXiv},
       eprint = {astro-ph/9403052},
 primaryClass = {astro-ph},
       adsurl = {https://ui.adsabs.harvard.edu/abs/1994ApJ...428L..13N}
}

@ARTICLE{UPP,
       author = {{Arnaud}, M. and {Pratt}, G.~W. and {Piffaretti}, R. and {B{\"o}hringer}, H. and {Croston}, J.~H. and {Pointecouteau}, E.},
        title = "{The universal galaxy cluster pressure profile from a representative sample of nearby systems (REXCESS) and the Y$_{SZ}$ - M$_{500}$ relation}",
      journal = {\aap},
         year = 2010,
        month = jul,
       volume = {517},
          eid = {A92},
        pages = {A92},
          doi = {10.1051/0004-6361/200913416},
archivePrefix = {arXiv},
       eprint = {0910.1234},
 primaryClass = {astro-ph.CO},
       adsurl = {https://ui.adsabs.harvard.edu/abs/2010A&A...517A..92A}
}

@ARTICLE{Bleem2022,
       author = {{Bleem}, L.~E. and {Crawford}, T.~M. and {Ansarinejad}, B. and {Benson}, B.~A. and {Bocquet}, S. and {Carlstrom}, J.~E. and {Chang}, C.~L. and {Chown}, R. and {Crites}, A.~T. and {de Haan}, T. and {Dobbs}, M.~A. and {Everett}, W.~B. and {George}, E.~M. and {Gualtieri}, R. and {Halverson}, N.~W. and {Holder}, G.~P. and {Holzapfel}, W.~L. and {Hrubes}, J.~D. and {Knox}, L. and {Lee}, A.~T. and {Luong-Van}, D. and {Marrone}, D.~P. and {McMahon}, J.~J. and {Meyer}, S.~S. and {Millea}, M. and {Mocanu}, L.~M. and {Mohr}, J.~J. and {Natoli}, T. and {Omori}, Y. and {Padin}, S. and {Pryke}, C. and {Raghunathan}, S. and {Reichardt}, C.~L. and {Ruhl}, J.~E. and {Schaffer}, K.~K. and {Shirokoff}, E. and {Staniszewski}, Z. and {Stark}, A.~A. and {Vieira}, J.~D. and {Williamson}, R.},
        title = "{CMB/kSZ and Compton-y Maps from 2500 deg$^{2}$ of SPT-SZ and Planck Survey Data}",
      journal = {\apjs},
         year = 2022,
        month = feb,
       volume = {258},
       number = {2},
          eid = {36},
        pages = {36},
          doi = {10.3847/1538-4365/ac35e9},
archivePrefix = {arXiv},
       eprint = {2102.05033},
 primaryClass = {astro-ph.CO},
       adsurl = {https://ui.adsabs.harvard.edu/abs/2022ApJS..258...36B}
}

@ARTICLE{Shirasaki2019,
       author = {{Shirasaki}, Masato},
        title = "{Impact of radio sources and cosmic infrared background on thermal Sunyaev-Zel'dovich - gravitational lensing cross-correlation}",
      journal = {\mnras},
         year = 2019,
        month = feb,
       volume = {483},
       number = {1},
        pages = {342-351},
          doi = {10.1093/mnras/sty3162},
archivePrefix = {arXiv},
       eprint = {1807.09412},
 primaryClass = {astro-ph.CO},
       adsurl = {https://ui.adsabs.harvard.edu/abs/2019MNRAS.483..342S}
}

@ARTICLE{Raghunathan2023,
       author = {{Raghunathan}, Srinivasan and {Omori}, Yuuki},
        title = "{A Cross-internal Linear Combination Approach to Probe the Secondary CMB Anisotropies: Kinematic Sunyaev-Zel'dovich Effect and CMB Lensing}",
      journal = {\apj},
         year = 2023,
        month = sep,
       volume = {954},
       number = {1},
          eid = {83},
        pages = {83},
          doi = {10.3847/1538-4357/ace0c6},
archivePrefix = {arXiv},
       eprint = {2304.09166},
 primaryClass = {astro-ph.CO},
       adsurl = {https://ui.adsabs.harvard.edu/abs/2023ApJ...954...83R}
}

@ARTICLE{Dicker2021,
       author = {{Dicker}, Simon R. and {Battistelli}, Elia S. and {Bhandarkar}, Tanay and {Devlin}, Mark J. and {Duff}, Shannon M. and {Hilton}, Gene and {Hilton}, Matt and {Hincks}, Adam D. and {Hubmayr}, Johannes and {Huffenberger}, Kevin and {Hughes}, John P. and {Di Mascolo}, Luca and {Mason}, Brian S. and {Mates}, J.~A.~B. and {McMahon}, Jeff and {Mroczkowski}, Tony and {Naess}, Sigurd and {Orlowski-Scherer}, John and {Partridge}, Bruce and {Radiconi}, Federico and {Romero}, Charles and {Sarazin}, Craig L. and {Sehgal}, Neelima and {Sievers}, Jonathan and {Sif{\'o}n}, Crist{\'o}bal and {Ullom}, Joel and {Vale}, Leila R. and {Vissers}, Michael R. and {Xu}, Zhilei},
        title = "{Observations of compact sources in galaxy clusters using MUSTANG2}",
      journal = {\mnras},
         year = 2021,
        month = dec,
       volume = {508},
       number = {2},
        pages = {2600-2612},
          doi = {10.1093/mnras/stab2679},
archivePrefix = {arXiv},
       eprint = {2107.06725},
 primaryClass = {astro-ph.CO},
       adsurl = {https://ui.adsabs.harvard.edu/abs/2021MNRAS.508.2600D}
}

@ARTICLE{Chluba2012,
       author = {{Chluba}, Jens and {Nagai}, Daisuke and {Sazonov}, Sergey and {Nelson}, Kaylea},
        title = "{A fast and accurate method for computing the Sunyaev-Zel'dovich signal of hot galaxy clusters}",
      journal = {\mnras},
         year = 2012,
        month = oct,
       volume = {426},
       number = {1},
        pages = {510-530},
          doi = {10.1111/j.1365-2966.2012.21741.x},
archivePrefix = {arXiv},
       eprint = {1205.5778},
 primaryClass = {astro-ph.CO},
       adsurl = {https://ui.adsabs.harvard.edu/abs/2012MNRAS.426..510C}
}

@ARTICLE{Itoh1998,
       author = {{Itoh}, Naoki and {Kohyama}, Yasuharu and {Nozawa}, Satoshi},
        title = "{Relativistic Corrections to the Sunyaev-Zeldovich Effect for Clusters of Galaxies}",
      journal = {\apj},
         year = 1998,
        month = jul,
       volume = {502},
       number = {1},
        pages = {7-15},
          doi = {10.1086/305876},
archivePrefix = {arXiv},
       eprint = {astro-ph/9712289},
 primaryClass = {astro-ph},
       adsurl = {https://ui.adsabs.harvard.edu/abs/1998ApJ...502....7I}
}

@ARTICLE{1970Ap&SS...7....3S,
       author = {{Sunyaev}, R.~A. and {Zeldovich}, Ya. B.},
        title = "{Small-Scale Fluctuations of Relic Radiation}",
      journal = {\apss},
         year = 1970,
        month = apr,
       volume = {7},
       number = {1},
        pages = {3-19},
          doi = {10.1007/BF00653471},
       adsurl = {https://ui.adsabs.harvard.edu/abs/1970Ap&SS...7....3S}
}

@ARTICLE{Erosita2022,
       author = {{Comparat}, Johan and {Truong}, Nhut and {Merloni}, Andrea and {Pillepich}, Annalisa and {Ponti}, Gabriele and {Driver}, Simon and {Bellstedt}, Sabine and {Liske}, Joe and {Aird}, James and {Br{\"u}ggen}, Marcus and {Bulbul}, Esra and {Davies}, Luke and {Villalba}, Justo Antonio Gonz{\'a}lez and {Georgakakis}, Antonis and {Haberl}, Frank and {Liu}, Teng and {Maitra}, Chandreyee and {Nandra}, Kirpal and {Popesso}, Paola and {Predehl}, Peter and {Robotham}, Aaron and {Salvato}, Mara and {Thorne}, Jessica E. and {Zhang}, Yi},
        title = "{The eROSITA Final Equatorial Depth Survey (eFEDS). X-ray emission around star-forming and quiescent galaxies at 0.05 < z < 0.3}",
      journal = {\aap},
         year = 2022,
        month = oct,
       volume = {666},
          eid = {A156},
        pages = {A156},
          doi = {10.1051/0004-6361/202243101},
archivePrefix = {arXiv},
       eprint = {2201.05169},
 primaryClass = {astro-ph.GA},
       adsurl = {https://ui.adsabs.harvard.edu/abs/2022A&A...666A.156C}
}

@ARTICLE{Erosita2024,
       author = {{Zhang}, Yi and {Comparat}, Johan and {Ponti}, Gabriele and {Merloni}, Andrea and {Nandra}, Kirpal and {Haberl}, Frank and {Locatelli}, Nicola and {Zhang}, Xiaoyuan and {Sanders}, Jeremy and {Zheng}, Xueying and {Liu}, Ang and {Popesso}, Paola and {Liu}, Teng and {Truong}, Nhut and {Pillepich}, Annalisa and {Predehl}, Peter and {Salvato}, Mara and {Shreeram}, Soumya and {Yeung}, Michael C.~H. and {Ni}, Qingling},
        title = "{The hot circumgalactic medium in the eROSITA All-Sky Survey: I. X-ray surface brightness profiles}",
      journal = {\aap},
         year = 2024,
        month = oct,
       volume = {690},
          eid = {A267},
        pages = {A267},
          doi = {10.1051/0004-6361/202449412},
archivePrefix = {arXiv},
       eprint = {2401.17308},
 primaryClass = {astro-ph.GA},
       adsurl = {https://ui.adsabs.harvard.edu/abs/2024A&A...690A.267Z}
}

@ARTICLE{Tumlinson2017,
       author = {{Tumlinson}, Jason and {Peeples}, Molly S. and {Werk}, Jessica K.},
        title = "{The Circumgalactic Medium}",
      journal = {\araa},
         year = 2017,
        month = aug,
       volume = {55},
       number = {1},
        pages = {389-432},
          doi = {10.1146/annurev-astro-091916-055240},
archivePrefix = {arXiv},
       eprint = {1709.09180},
 primaryClass = {astro-ph.GA},
       adsurl = {https://ui.adsabs.harvard.edu/abs/2017ARA&A..55..389T}
}

@BOOK{Astro2020,
  author    = "{National Academies of Sciences, Engineering, and Medicine}",
  title     = "Pathways to Discovery in Astronomy and Astrophysics for the 2020s",
  isbn      = "978-0-309-46734-6",
  doi       = "10.17226/26141",
  url       = "https://nap.nationalacademies.org/catalog/26141/pathways-to-discovery-in-astronomy-and-astrophysics-for-the-2020s",
  year      = 2023,
  publisher = "The National Academies Press",
  address   = "Washington, DC"
  }

@ARTICLE{NVSS,
       author = {{Condon}, J.~J. and {Cotton}, W.~D. and {Greisen}, E.~W. and {Yin}, Q.~F. and {Perley}, R.~A. and {Broderick}, J.~J.},
        title = "{The NRAO VLA Sky Survey}",
      journal = {Astronomy Data Image Library},
         year = 1996,
        month = mar,
       adsurl = {https://ui.adsabs.harvard.edu/abs/1996ADIL...JC...01C}
}

@ARTICLE{Mandelbaum2008,
       author = {{Mandelbaum}, Rachel and {Li}, Cheng and {Kauffmann}, Guinevere and {White}, Simon D.~M.},
        title = "{Halo masses for optically selected and for radio-loud AGN from clustering and galaxy-galaxy lensing}",
      journal = {\mnras},
         year = 2009,
        month = feb,
       volume = {393},
       number = {2},
        pages = {377-392},
          doi = {10.1111/j.1365-2966.2008.14235.x},
archivePrefix = {arXiv},
       eprint = {0806.4089},
 primaryClass = {astro-ph},
       adsurl = {https://ui.adsabs.harvard.edu/abs/2009MNRAS.393..377M}
}

@ARTICLE{LeBrun2014,
       author = {{Le Brun}, Amandine M.~C. and {McCarthy}, Ian G. and {Melin}, Jean-Baptiste},
        title = "{Testing Sunyaev-Zel'dovich measurements of the hot gas content of dark matter haloes using synthetic skies}",
      journal = {\mnras},
         year = 2015,
        month = aug,
       volume = {451},
       number = {4},
        pages = {3868-3881},
          doi = {10.1093/mnras/stv1172},
archivePrefix = {arXiv},
       eprint = {1501.05666},
 primaryClass = {astro-ph.CO},
       adsurl = {https://ui.adsabs.harvard.edu/abs/2015MNRAS.451.3868L}
}

@ARTICLE{Hall2019,
       author = {{Hall}, Kirsten R. and {Zakamska}, Nadia L. and {Addison}, Graeme E. and {Battaglia}, Nicholas and {Crichton}, Devin and {Devlin}, Mark and {Dunkley}, Joanna and {Gralla}, Megan and {Hill}, J. Colin and {Hilton}, Matt and {Hubmayr}, Johannes and {Hughes}, John P. and {Huffenberger}, Kevin M. and {Kosowsky}, Arthur and {Marriage}, Tobias A. and {Maurin}, Lo{\"\i}c and {Moodley}, Kavilan and {Niemack}, Michael D. and {Page}, Lyman A. and {Partridge}, Bruce and {D{\"u}nner Planella}, Rolando and {Schillaci}, Alessandro and {Sif{\'o}n}, Crist{\'o}bal and {Staggs}, Suzanne T. and {Wollack}, Edward J. and {Xu}, Zhilei},
        title = "{Quantifying the thermal Sunyaev-Zel'dovich effect and excess millimetre emission in quasar environments}",
      journal = {\mnras},
         year = 2019,
        month = dec,
       volume = {490},
       number = {2},
        pages = {2315-2335},
          doi = {10.1093/mnras/stz2751},
archivePrefix = {arXiv},
       eprint = {1907.11731},
 primaryClass = {astro-ph.GA},
       adsurl = {https://ui.adsabs.harvard.edu/abs/2019MNRAS.490.2315H}
}

@INPROCEEDINGS{DSA2000,
       author = {{Hallinan}, Gregg and {Ravi}, V. and {Weinreb}, S. and {Kocz}, J. and {Huang}, Y. and {Woody}, D.~P. and {Lamb}, J. and {D'Addario}, L. and {Catha}, M. and {Law}, C. and {Kulkarni}, S.~R. and {Phinney}, E.~S. and {Eastwood}, M.~W. and {Bouman}, K. and {McLaughlin}, M. and {Ransom}, S. and {Siemens}, X. and {Cordes}, J. and {Lynch}, R. and {Kaplan}, D. and {Brazier}, A. and {Bhatnagar}, S. and {Myers}, S. and {Walter}, F. and {Gaensler}, B.},
        title = "{The DSA-2000 {\textemdash} A Radio Survey Camera}",
    booktitle = {Bulletin of the American Astronomical Society},
         year = 2019,
       volume = {51},
        month = sep,
          eid = {255},
        pages = {255},
          doi = {10.48550/arXiv.1907.07648},
archivePrefix = {arXiv},
       eprint = {1907.07648},
 primaryClass = {astro-ph.IM},
}

@ARTICLE{First,
       author = {{Helfand}, David J. and {White}, Richard L. and {Becker}, Robert H.},
        title = "{The Last of FIRST: The Final Catalog and Source Identifications}",
      journal = {\apj},
         year = 2015,
        month = mar,
       volume = {801},
       number = {1},
          eid = {26},
        pages = {26},
          doi = {10.1088/0004-637X/801/1/26},
archivePrefix = {arXiv},
       eprint = {1501.01555},
 primaryClass = {astro-ph.GA},
       adsurl = {https://ui.adsabs.harvard.edu/abs/2015ApJ...801...26H}
}

@ARTICLE{Coulton2026,
       author = {{Coulton}, William R. and {Duivenvoorden}, Adriaan J. and {Atkins}, Zachary and {Battaglia}, Nicholas and {Battistelli}, Elia Stefano and {Bond}, J. Richard and {Cai}, Hongbo and {Calabrese}, Erminia and {Choi}, Steve K. and {Crowley}, Kevin T. and {Devlin}, Mark J. and {Dunkley}, Jo and {Ferraro}, Simone and {Guan}, Yilun and {Herv{\'\i}as-Caimapo}, Carlos and {Hill}, J. Colin and {Hilton}, Matt and {Hincks}, Adam D. and {Kosowsky}, Arthur and {Madhavacheril}, Mathew S. and {van Marrewijk}, Joshiwa and {McCarthy}, Fiona and {Moodley}, Kavilan and {Mroczkowski}, Tony and {Niemack}, Michael D. and {Page}, Lyman A. and {Partridge}, Bruce and {Schaan}, Emmanuel and {Sehgal}, Neelima and {Sherwin}, Blake D. and {Sif{\'o}n}, Crist{\'o}bal and {Spergel}, David N. and {Staggs}, Suzanne T. and {Van Engelen}, Alexander and {Vavagiakis}, Eve M. and {Wollack}, Edward J.},
        title = "{Atacama Cosmology Telescope: A measurement of galaxy cluster temperatures through relativistic corrections to the thermal Sunyaev-Zeldovich effect}",
      journal = {\prd},
         year = 2026,
        month = feb,
       volume = {113},
       number = {4},
          eid = {043520},
        pages = {043520},
          doi = {10.1103/n7p5-pc66},
archivePrefix = {arXiv},
       eprint = {2410.19046},
 primaryClass = {astro-ph.CO},
       adsurl = {https://ui.adsabs.harvard.edu/abs/2026PhRvD.113d3520C}
}

@ARTICLE{Koukoufilippas2020,
       author = {{Koukoufilippas}, Nick and {Alonso}, David and {Bilicki}, Maciej and {Peacock}, John A.},
        title = "{Tomographic measurement of the intergalactic gas pressure through galaxy-tSZ cross-correlations}",
      journal = {\mnras},
         year = 2020,
        month = feb,
       volume = {491},
       number = {4},
        pages = {5464-5480},
          doi = {10.1093/mnras/stz3351},
archivePrefix = {arXiv},
       eprint = {1909.09102},
 primaryClass = {astro-ph.CO},
       adsurl = {https://ui.adsabs.harvard.edu/abs/2020MNRAS.491.5464K}
}

@ARTICLE{Dicker2024,
       author = {{Dicker}, Simon R. and {Perez Sarmiento}, Karen and {Mason}, Brian and {Bhandarkar}, Tanay and {Devlin}, Mark J. and {Di Mascolo}, Luca and {Haridas}, Saianeesh and {Hilton}, Matt and {Madhavacheril}, Mathew and {Moravec}, Emily and {Mroczkowski}, Tony and {Orlowski-Scherer}, John and {Romero}, Charles and {Sarazin}, Craig L. and {Sievers}, Jonathan},
        title = "{Sensitive 3 mm Imaging of Discrete Sources in the Fields of Thermal Sunyaev─Zel'dovich Effect─Selected Galaxy Clusters}",
      journal = {\apj},
         year = 2024,
        month = jul,
       volume = {970},
       number = {1},
          eid = {84},
        pages = {84},
          doi = {10.3847/1538-4357/ad4e35},
archivePrefix = {arXiv},
       eprint = {2403.09855},
 primaryClass = {astro-ph.CO},
       adsurl = {https://ui.adsabs.harvard.edu/abs/2024ApJ...970...84D}
}

@ARTICLE{Sanchez2023,
       author = {{S{\'a}nchez}, J. and {Omori}, Y. and {Chang}, C. and {Bleem}, L.~E. and {Crawford}, T. and {Drlica-Wagner}, A. and {Raghunathan}, S. and {Zacharegkas}, G. and {Abbott}, T.~M.~C. and {Aguena}, M. and {Alarcon}, A. and {Allam}, S. and {Alves}, O. and {Amon}, A. and {Avila}, S. and {Baxter}, E. and {Bechtol}, K. and {Benson}, B.~A. and {Bernstein}, G.~M. and {Bertin}, E. and {Bocquet}, S. and {Brooks}, D. and {Burke}, D.~L. and {Campos}, A. and {Carlstrom}, J.~E. and {Rosell}, A. Carnero and {Kind}, M. Carrasco and {Carretero}, J. and {Castander}, F.~J. and {Cawthon}, R. and {Chang}, C.~L. and {Chen}, A. and {Choi}, A. and {Chown}, R. and {Costanzi}, M. and {Crites}, A.~T. and {Crocce}, M. and {da Costa}, L.~N. and {Pereira}, M.~E.~S. and {de Haan}, T. and {De Vicente}, J. and {DeRose}, J. and {Desai}, S. and {Diehl}, H.~T. and {Dobbs}, M.~A. and {Dodelson}, S. and {Doel}, P. and {Elvin-Poole}, J. and {Everett}, W. and {Everett}, S. and {Ferrero}, I. and {Flaugher}, B. and {Fosalba}, P. and {Frieman}, J. and {Garc{\'\i}a-Bellido}, J. and {Gatti}, M. and {George}, E.~M. and {Gerdes}, D.~W. and {Giannini}, G. and {Gruen}, D. and {Gruendl}, R.~A. and {Gschwend}, J. and {Gutierrez}, G. and {Halverson}, N.~W. and {Hinton}, S.~R. and {Holder}, G.~P. and {Hollowood}, D.~L. and {Holzapfel}, W.~L. and {Honscheid}, K. and {Hrubes}, J.~D. and {James}, D.~J. and {Knox}, L. and {Kuehn}, K. and {Kuropatkin}, N. and {Lahav}, O. and {Lee}, A.~T. and {Luong-Van}, D. and {MacCrann}, N. and {Marshall}, J.~L. and {McCullough}, J. and {McMahon}, J.~J. and {Melchior}, P. and {Mena-Fern{\'a}ndez}, J. and {Menanteau}, F. and {Miquel}, R. and {Mocanu}, L. and {Mohr}, J.~J. and {Muir}, J. and {Myles}, J. and {Natoli}, T. and {Padin}, S. and {Palmese}, A. and {Pandey}, S. and {Paz-Chinch{\'o}n}, F. and {Pieres}, A. and {Malag{\'o}n}, A.~A. Plazas and {Porredon}, A. and {Pryke}, C. and {Raveri}, M. and {Reichardt}, C.~L. and {Rodriguez-Monroy}, M. and {Ross}, A.~J. and {Ruhl}, J.~E. and {Rykoff}, E. and {S{\'a}nchez}, C. and {Sanchez}, E. and {Scarpine}, V. and {Schaffer}, K.~K. and {Sevilla-Noarbe}, I. and {Sheldon}, E. and {Shirokoff}, E. and {Smith}, M. and {Soares-Santos}, M. and {Staniszewski}, Z. and {Stark}, A.~A. and {Suchyta}, E. and {Swanson}, M.~E.~C. and {Tarle}, G. and {Thomas}, D. and {Troxel}, M.~A. and {Tucker}, D.~L. and {Vieira}, J.~D. and {Vincenzi}, M. and {Weaverdyck}, N. and {Williamson}, R. and {Yanny}, B. and {Yin}, B. and {DES Collaboration} and {SPT Collaboration}},
        title = "{Mapping gas around massive galaxies: cross-correlation of DES Y3 galaxies and Compton-y maps from SPT and Planck}",
      journal = {\mnras},
         year = 2023,
        month = jun,
       volume = {522},
       number = {2},
        pages = {3163-3182},
          doi = {10.1093/mnras/stad1167},
archivePrefix = {arXiv},
       eprint = {2210.08633},
 primaryClass = {astro-ph.CO},
       adsurl = {https://ui.adsabs.harvard.edu/abs/2023MNRAS.522.3163S}
}

@ARTICLE{BFSS2017,
       author = {{Battaglia}, Nicholas and {Ferraro}, Simone and {Schaan}, Emmanuel and {Spergel}, David N.},
        title = "{Future constraints on halo thermodynamics from combined Sunyaev-Zel'dovich measurements}",
      journal = {\jcap},
         year = 2017,
        month = nov,
       volume = {2017},
       number = {11},
          eid = {040},
        pages = {040},
          doi = {10.1088/1475-7516/2017/11/040},
archivePrefix = {arXiv},
       eprint = {1705.05881},
 primaryClass = {astro-ph.CO},
       adsurl = {https://ui.adsabs.harvard.edu/abs/2017JCAP...11..040B}
}

@ARTICLE{Hand2011,
       author = {{Hand}, Nick and {Appel}, John W. and {Battaglia}, Nick and {Bond}, J. Richard and {Das}, Sudeep and {Devlin}, Mark J. and {Dunkley}, Joanna and {D{\"u}nner}, Rolando and {Essinger-Hileman}, Thomas and {Fowler}, Joseph W. and {Hajian}, Amir and {Halpern}, Mark and {Hasselfield}, Matthew and {Hilton}, Matt and {Hincks}, Adam D. and {Hlozek}, Ren{\'e}e and {Hughes}, John P. and {Irwin}, Kent D. and {Klein}, Jeff and {Kosowsky}, Arthur and {Lin}, Yen-Ting and {Marriage}, Tobias A. and {Marsden}, Danica and {McLaren}, Mike and {Menanteau}, Felipe and {Moodley}, Kavilan and {Niemack}, Michael D. and {Nolta}, Michael R. and {Page}, Lyman A. and {Parker}, Lucas and {Partridge}, Bruce and {Plimpton}, Reed and {Reese}, Erik D. and {Rojas}, Felipe and {Sehgal}, Neelima and {Sherwin}, Blake D. and {Sievers}, Jonathan L. and {Spergel}, David N. and {Staggs}, Suzanne T. and {Swetz}, Daniel S. and {Switzer}, Eric R. and {Thornton}, Robert and {Trac}, Hy and {Visnjic}, Katerina and {Wollack}, Ed},
        title = "{The Atacama Cosmology Telescope: Detection of Sunyaev-Zel'Dovich Decrement in Groups and Clusters Associated with Luminous Red Galaxies}",
      journal = {\apj},
         year = 2011,
        month = jul,
       volume = {736},
       number = {1},
          eid = {39},
        pages = {39},
          doi = {10.1088/0004-637X/736/1/39},
archivePrefix = {arXiv},
       eprint = {1101.1951},
 primaryClass = {astro-ph.CO},
       adsurl = {https://ui.adsabs.harvard.edu/abs/2011ApJ...736...39H}
}

@ARTICLE{Spacek2016,
       author = {{Spacek}, Alexander and {Scannapieco}, Evan and {Cohen}, Seth and {Joshi}, Bhavin and {Mauskopf}, Philip},
        title = "{Constraining AGN Feedback in Massive Ellipticals with South Pole Telescope Measurements of the Thermal Sunyaev-Zel'dovich Effect}",
      journal = {\apj},
         year = 2016,
        month = mar,
       volume = {819},
       number = {2},
          eid = {128},
        pages = {128},
          doi = {10.3847/0004-637X/819/2/128},
archivePrefix = {arXiv},
       eprint = {1601.01330},
 primaryClass = {astro-ph.GA},
       adsurl = {https://ui.adsabs.harvard.edu/abs/2016ApJ...819..128S}
}

@ARTICLE{Soergel2017,
       author = {{Soergel}, Bjoern and {Giannantonio}, Tommaso and {Efstathiou}, George and {Puchwein}, Ewald and {Sijacki}, Debora},
        title = "{Constraints on AGN feedback from its Sunyaev-Zel'dovich imprint on the cosmic background radiation}",
      journal = {\mnras},
         year = 2017,
        month = jun,
       volume = {468},
       number = {1},
        pages = {577-596},
          doi = {10.1093/mnras/stx492},
archivePrefix = {arXiv},
       eprint = {1612.06296},
 primaryClass = {astro-ph.CO},
       adsurl = {https://ui.adsabs.harvard.edu/abs/2017MNRAS.468..577S}
}

@ARTICLE{McCarthy2014,
       author = {{McCarthy}, I.~G. and {Le Brun}, A.~M.~C. and {Schaye}, J. and {Holder}, G.~P.},
        title = "{The thermal Sunyaev-Zel'dovich effect power spectrum in light of Planck}",
      journal = {\mnras},
         year = 2014,
        month = jun,
       volume = {440},
       number = {4},
        pages = {3645-3657},
          doi = {10.1093/mnras/stu543},
archivePrefix = {arXiv},
       eprint = {1312.5341},
 primaryClass = {astro-ph.CO},
       adsurl = {https://ui.adsabs.harvard.edu/abs/2014MNRAS.440.3645M}
}

@ARTICLE{BBPSS,
       author = {{Battaglia}, N. and {Bond}, J.~R. and {Pfrommer}, C. and {Sievers}, J.~L. and {Sijacki}, D.},
        title = "{Simulations of the Sunyaev-Zel'dovich Power Spectrum with Active Galactic Nucleus Feedback}",
      journal = {\apj},
         year = 2010,
        month = dec,
       volume = {725},
       number = {1},
        pages = {91-99},
          doi = {10.1088/0004-637X/725/1/91},
archivePrefix = {arXiv},
       eprint = {1003.4256},
 primaryClass = {astro-ph.CO},
       adsurl = {https://ui.adsabs.harvard.edu/abs/2010ApJ...725...91B}
}

@ARTICLE{Meinke2021,
       author = {{Meinke}, Jeremy and {B{\"o}ckmann}, Kathrin and {Cohen}, Seth and {Mauskopf}, Philip and {Scannapieco}, Evan and {Sarmento}, Richard and {Lunde}, Emily and {Cottle}, J'Neil},
        title = "{The Thermal Sunyaev-Zel'dovich Effect from Massive, Quiescent 0.5 {\ensuremath{\leq}} z {\ensuremath{\leq}} 1.5 Galaxies}",
      journal = {\apj},
         year = 2021,
        month = jun,
       volume = {913},
       number = {2},
          eid = {88},
        pages = {88},
          doi = {10.3847/1538-4357/abf2b4},
archivePrefix = {arXiv},
       eprint = {2103.01245},
 primaryClass = {astro-ph.CO},
       adsurl = {https://ui.adsabs.harvard.edu/abs/2021ApJ...913...88M}
}

@ARTICLE{Scannapieco2008,
       author = {{Scannapieco}, Evan and {Thacker}, Robert J. and {Couchman}, H.~M.~P.},
        title = "{Measuring AGN Feedback with the Sunyaev-Zel'dovich Effect}",
      journal = {\apj},
         year = 2008,
        month = may,
       volume = {678},
       number = {2},
        pages = {674-685},
          doi = {10.1086/528948},
archivePrefix = {arXiv},
       eprint = {0709.0952},
 primaryClass = {astro-ph},
       adsurl = {https://ui.adsabs.harvard.edu/abs/2008ApJ...678..674S}
}

@ARTICLE{Grayson2023,
       author = {{Grayson}, Skylar and {Scannapieco}, Evan and {Dav{\'e}}, Romeel},
        title = "{Distinguishing Active Galactic Nuclei Feedback Models with the Thermal Sunyaev-Zel'dovich Effect}",
      journal = {\apj},
         year = 2023,
        month = nov,
       volume = {957},
       number = {1},
          eid = {17},
        pages = {17},
          doi = {10.3847/1538-4357/acfd26},
archivePrefix = {arXiv},
       eprint = {2310.01502},
 primaryClass = {astro-ph.GA},
       adsurl = {https://ui.adsabs.harvard.edu/abs/2023ApJ...957...17G}
}

@ARTICLE{Hu2007,
       author = {{Hu}, Wayne and {DeDeo}, Simon and {Vale}, Chris},
        title = "{Cluster mass estimators from CMB temperature and polarization lensing}",
      journal = {New Journal of Physics},
         year = 2007,
        month = dec,
       volume = {9},
       number = {12},
        pages = {441},
          doi = {10.1088/1367-2630/9/12/441},
archivePrefix = {arXiv},
       eprint = {astro-ph/0701276},
 primaryClass = {astro-ph},
       adsurl = {https://ui.adsabs.harvard.edu/abs/2007NJPh....9..441H}
}

@ARTICLE{Madhavacheril2024,
       author = {{Madhavacheril}, Mathew S. and {Qu}, Frank J. and {Sherwin}, Blake D. and {MacCrann}, Niall and {Li}, Yaqiong and {Abril-Cabezas}, Irene and {Ade}, Peter A.~R. and {Aiola}, Simone and {Alford}, Tommy and {Amiri}, Mandana and {Amodeo}, Stefania and {An}, Rui and {Atkins}, Zachary and {Austermann}, Jason E. and {Battaglia}, Nicholas and {Battistelli}, Elia Stefano and {Beall}, James A. and {Bean}, Rachel and {Beringue}, Benjamin and {Bhandarkar}, Tanay and {Biermann}, Emily and {Bolliet}, Boris and {Bond}, J. Richard and {Cai}, Hongbo and {Calabrese}, Erminia and {Calafut}, Victoria and {Capalbo}, Valentina and {Carrero}, Felipe and {Challinor}, Anthony and {Chesmore}, Grace E. and {Cho}, Hsiao-mei and {Choi}, Steve K. and {Clark}, Susan E. and {C{\'o}rdova Rosado}, Rodrigo and {Cothard}, Nicholas F. and {Coughlin}, Kevin and {Coulton}, William and {Crowley}, Kevin T. and {Dalal}, Roohi and {Darwish}, Omar and {Devlin}, Mark J. and {Dicker}, Simon and {Doze}, Peter and {Duell}, Cody J. and {Duff}, Shannon M. and {Duivenvoorden}, Adriaan J. and {Dunkley}, Jo and {D{\"u}nner}, Rolando and {Fanfani}, Valentina and {Fankhanel}, Max and {Farren}, Gerrit and {Ferraro}, Simone and {Freundt}, Rodrigo and {Fuzia}, Brittany and {Gallardo}, Patricio A. and {Garrido}, Xavier and {Givans}, Jahmour and {Gluscevic}, Vera and {Golec}, Joseph E. and {Guan}, Yilun and {Hall}, Kirsten R. and {Halpern}, Mark and {Han}, Dongwon and {Harrison}, Ian and {Hasselfield}, Matthew and {Healy}, Erin and {Henderson}, Shawn and {Hensley}, Brandon and {Herv{\'\i}as-Caimapo}, Carlos and {Hill}, J. Colin and {Hilton}, Gene C. and {Hilton}, Matt and {Hincks}, Adam D. and {Hlo{\v{z}}ek}, Ren{\'e}e and {Ho}, Shuay-Pwu Patty and {Huber}, Zachary B. and {Hubmayr}, Johannes and {Huffenberger}, Kevin M. and {Hughes}, John P. and {Irwin}, Kent and {Isopi}, Giovanni and {Jense}, Hidde T. and {Keller}, Ben and {Kim}, Joshua and {Knowles}, Kenda and {Koopman}, Brian J. and {Kosowsky}, Arthur and {Kramer}, Darby and {Kusiak}, Aleksandra and {La Posta}, Adrien and {Lague}, Alex and {Lakey}, Victoria and {Lee}, Eunseong and {Li}, Zack and {Limon}, Michele and {Lokken}, Martine and {Louis}, Thibaut and {Lungu}, Marius and {MacInnis}, Amanda and {Maldonado}, Diego and {Maldonado}, Felipe and {Mallaby-Kay}, Maya and {Marques}, Gabriela A. and {McMahon}, Jeff and {Mehta}, Yogesh and {Menanteau}, Felipe and {Moodley}, Kavilan and {Morris}, Thomas W. and {Mroczkowski}, Tony and {Naess}, Sigurd and {Namikawa}, Toshiya and {Nati}, Federico and {Newburgh}, Laura and {Nicola}, Andrina and {Niemack}, Michael D. and {Nolta}, Michael R. and {Orlowski-Scherer}, John and {Page}, Lyman A. and {Pandey}, Shivam and {Partridge}, Bruce and {Prince}, Heather and {Puddu}, Roberto and {Radiconi}, Federico and {Robertson}, Naomi and {Rojas}, Felipe and {Sakuma}, Tai and {Salatino}, Maria and {Schaan}, Emmanuel and {Schmitt}, Benjamin L. and {Sehgal}, Neelima and {Shaikh}, Shabbir and {Sierra}, Carlos and {Sievers}, Jon and {Sif{\'o}n}, Crist{\'o}bal and {Simon}, Sara and {Sonka}, Rita and {Spergel}, David N. and {Staggs}, Suzanne T. and {Storer}, Emilie and {Switzer}, Eric R. and {Tampier}, Niklas and {Thornton}, Robert and {Trac}, Hy and {Treu}, Jesse and {Tucker}, Carole and {Ullom}, Joel and {Vale}, Leila R. and {Van Engelen}, Alexander and {Van Lanen}, Jeff and {van Marrewijk}, Joshiwa and {Vargas}, Cristian and {Vavagiakis}, Eve M. and {Wagoner}, Kasey and {Wang}, Yuhan and {Wenzl}, Lukas and {Wollack}, Edward J. and {Xu}, Zhilei and {Zago}, Fernando and {Zheng}, Kaiwen},
        title = "{The Atacama Cosmology Telescope: DR6 Gravitational Lensing Map and Cosmological Parameters}",
      journal = {\apj},
         year = 2024,
        month = feb,
       volume = {962},
       number = {2},
          eid = {113},
        pages = {113},
          doi = {10.3847/1538-4357/acff5f},
archivePrefix = {arXiv},
       eprint = {2304.05203},
 primaryClass = {astro-ph.CO},
       adsurl = {https://ui.adsabs.harvard.edu/abs/2024ApJ...962..113M}
}

@ARTICLE{KiDS,
       author = {{de Jong}, Jelte T.~A. and {Verdoes Kleijn}, Gijs A. and {Kuijken}, Konrad H. and {Valentijn}, Edwin A.},
        title = "{The Kilo-Degree Survey}",
      journal = {Experimental Astronomy},
         year = 2013,
        month = jan,
       volume = {35},
       number = {1-2},
        pages = {25-44},
          doi = {10.1007/s10686-012-9306-1},
archivePrefix = {arXiv},
       eprint = {1206.1254},
 primaryClass = {astro-ph.CO},
       adsurl = {https://ui.adsabs.harvard.edu/abs/2013ExA....35...25D}
}

@ARTICLE{Simba,
       author = {{Dav{\'e}}, Romeel and {Angl{\'e}s-Alc{\'a}zar}, Daniel and {Narayanan}, Desika and {Li}, Qi and {Rafieferantsoa}, Mika H. and {Appleby}, Sarah},
        title = "{SIMBA: Cosmological simulations with black hole growth and feedback}",
      journal = {\mnras},
         year = 2019,
        month = jun,
       volume = {486},
       number = {2},
        pages = {2827-2849},
          doi = {10.1093/mnras/stz937},
archivePrefix = {arXiv},
       eprint = {1901.10203},
 primaryClass = {astro-ph.GA},
       adsurl = {https://ui.adsabs.harvard.edu/abs/2019MNRAS.486.2827D}
}

\end{document}